\documentclass{aastex631}
\usepackage{color}
\usepackage{url}    % can use \url{xxxxx} command.  when in a caption, use \protect\url{xxxxxx}

\newcommand{\Gaia}{{\sl Gaia}}
\newcommand{\TESS}{{\sl TESS}}

\newcommand{\Msun}{\mbox{$M_{\sun}$}}

\newcommand{\degree}{\mbox{$^{\circ}$}}

\newcommand{\etal}{et al.}
\newcommand{\eg}{e.g.}
\newcommand{\ie}{i.e.}

\newcommand{\chisq}{\mbox{$\chi^2$}}
\newcommand{\rchisq}{\mbox{$\tilde\chi^2$}}

\newcommand{\Ks}{\mbox{$K_S$}}
\newcommand{\degs}{\mbox{$^{\circ}$}}

\newcommand\TPS{$3\pi$~Survey}
\newcommand{\gps}{\ensuremath{g_{\rm P1}}}
\newcommand{\rps}{\ensuremath{r_{\rm P1}}}
\newcommand{\ips}{\ensuremath{i_{\rm P1}}}
\newcommand{\zps}{\ensuremath{z_{\rm P1}}}
\newcommand{\yps}{\ensuremath{y_{\rm P1}}}
\newcommand{\wps}{\ensuremath{w_{\rm P1}}}
\newcommand{\grizy}{\gps\,\rps\,\ips\,\zps\,\yps}

\newcommand{\WISE}{{\sl WISE}}
\newcommand{\NEOWISE}{{\sl NEOWISE}}

\newcommand{\intg}{\hbox{\sc int-g}}
\newcommand{\vlg}{\hbox{\sc vl-g}}

\newcommand{\ex}[2]{\mbox{#1$\times$10$^{#2}$}}  % don't use with AASTeX

\newcommand\note[1]{\textbf{\color{red}#1}}   % notes while writing draft
\newcommand\todo[1]{\textbf{\noindent \color{red} $\square$ #1}}
\newcommand\new[1]{{\color{blue}#1}}          % new text for referee
\renewcommand\note[1]{\hspace{-1sp}}             
\renewcommand\todo[1]{\hspace{-1sp}}             
\renewcommand\new[1]{\color{black}#1}

\newcommand{\targetlong}{2MASS~J06195260$-$2903592}
\newcommand{\target}{2MASS~J0619$-$2903}
\newcommand{\Js}{\mbox{$J_S$}}

\accepted{for publication in AJ, August 22, 2022}

\shorttitle{A Long-Lived Disk Around \target}
\shortauthors{Liu et al.}

\graphicspath{{./}{figures/}}

\begin{document}

\title{On The Unusual Variability of 2MASS~J06195260$-$2903592:\\
  A Long-Lived Disk around a Young Ultracool Dwarf}

\correspondingauthor{Michael C. Liu}
\email{mliu@ifa.hawaii.edu}

\author{Michael C. Liu}
%\author[0000-0003-2232-7664]{Michael C. Liu}
\affiliation{Institute for Astronomy, University of Hawai'i, 2680
  Woodlawn Drive, Honolulu HI 96822}
\affiliation{Visiting Astronomer at the Infrared Telescope Facility,
  which is operated by the University of Hawai'i under contract
  80HQTR19D0030 with the National Aeronautics and Space Administration.}

\author{Eugene A. Magnier}
\affiliation{Institute for Astronomy, University of Hawai'i, 2680
  Woodlawn Drive, Honolulu HI 96822}

\author{Zhoujian Zhang}
\affiliation{Visiting Astronomer at the Infrared Telescope Facility,
  which is operated by the University of Hawai'i under contract
  80HQTR19D0030 with the National Aeronautics and Space Administration.}
\affiliation{The University of Texas at Austin, Department of Astronomy,
  2515 Speedway C1400, Austin, TX 78712, USA}

\author{Eric Gaidos}
% ORCID  0000-0002-5258-684
\affiliation{Department of Earth Sciences, University of Hawai’i at
  Manoa, 1680 East-West Rd, Honolulu, HI 96822, USA}
\affiliation{Institute for Astronomy, University of Hawai'i, 2680
  Woodlawn Drive, Honolulu HI 96822}

\author{Trent J. Dupuy}
\affiliation{Institute for Astronomy, University of Edinburgh, Royal
  Observatory, Blackford Hill, Edinburgh, EH9 3HJ, UK}

\author{Pengyu Liu} \affiliation{SUPA, Institute for Astronomy, Royal
  Observatory, University of Edinburgh, Blackford Hill, Edinburgh
  EH93HJ, UK; Centre for Exoplanet Science, University of Edinburgh,
  Edinburgh EH9 3HJ, UK}

\author{Beth A. Biller}
\affiliation{SUPA, Institute for Astronomy, Royal Observatory,
  University of Edinburgh, Blackford Hill, Edinburgh EH93HJ, UK; Centre
  for Exoplanet Science, University of Edinburgh, Edinburgh EH9 3HJ, UK}

\author{Johanna M. Vos}
\affiliation{Department of Astrophysics, American
  Museum of Natural History, Central Park West at 79th Street, NY 10024,
  USA}

\author{Katelyn N. Allers}
\affiliation{Department of Physics and Astronomy, Bucknell University,
  Lewisburg, PA 17837}

\author{Jason T. Hinkle}
% ORCID: 0000-0001-9668-2920 
\affiliation{Institute for Astronomy, University of Hawai'i, 2680
  Woodlawn Drive, Honolulu HI 96822}

\author{Benjamin J. Shappee}
\affiliation{Institute for Astronomy, University of Hawai'i, 2680
  Woodlawn Drive, Honolulu HI 96822}

\author{Sage N. L. Constantinou}
\affiliation{Institute for Astronomy, University of Hawai'i, 2680
  Woodlawn Drive, Honolulu HI 96822}

\author{Mitchell T. Dennis}
\affiliation{Institute for Astronomy, University of Hawai'i, 2680
  Woodlawn Drive, Honolulu HI 96822}

\author{Kenji S. Emerson}
\affiliation{Institute for Astronomy, University of Hawai'i, 2680
  Woodlawn Drive, Honolulu HI 96822}

\begin{abstract}
  \noindent We present the characterization of the low-gravity M6
  dwarf \targetlong, previously identified as an unusual field object based on
  its strong IR excess and variable near-IR spectrum.
  Multiple epochs of low-resolution ($R\approx150$) near-IR spectra show
  large-amplitude ($\approx$0.1--0.5~mag) continuum variations on
  timescales of days to 12~years, unlike the small-amplitude variability
  typical for field ultracool dwarfs. The variations between epochs are
  well-modeled as changes in the relative extinction
  ($\Delta{A_V}\approx2$~mag).
  Similarly, Pan-STARRS\,1 optical photometry varies on timescales as
  long as 11~years (and possibly as short as an hour) and implies
  similarly amplitude of $A_V$ changes.
  \new{\NEOWISE\ mid-IR light curves also suggest changes on 6-month
    timescales, with amplitudes consistent with the optical/near-IR
    extinction variations. However, near-IR spectra, near-IR photometry,
    and optical photometry obtained in the past year indicate the source
    can also be stable on hourly and monthly timescales.}
  From comparison to objects of similar spectral type, the total
  extinction of \target\ seems to be $A_V\approx4-6$~mag, with perhaps
  epochs of lower extinction.
  \Gaia\ EDR3 finds that \target\ has a wide-separation (1\farcm2 =
  10450~AU) stellar companion, with an isochronal age of
  $31^{+22}_{-10}$\,Myr and a mass of $0.30^{+0.04}_{-0.03}$\,\Msun.
  Adopting this companion's age and EDR3 distance ($145.2\pm0.6$\,pc),
  we estimate a mass of 0.11--0.17~\Msun\ for \target. Altogether,
  \target\ appears to possess an unusually long-lived primordial
  circumstellar disk, perhaps making it a more obscured analog to the
  ``Peter Pan'' disks found around a few M~dwarfs in nearby young moving
  groups.
\end{abstract}


%----------------------------------------------------------------------%
\section{Introduction}

We present here characterization of the M6~dwarf \targetlong\
(hereinafter \target). This object was found by
\citet{2003AJ....126.2421C} as part of their 2MASS-based search for
ultracool members of the solar neighborhood. They noted that the
object's red optical spectrum showed signs of low gravity (\ie, youth),
based on weakness of the CaH (6750--7050~\AA) and \ion{K}{1}
$\lambda\lambda$ 7665, 7669~\AA\ absorption features relative to the M6
optical standard and similarity to the members of the TW~Hyd Association
(10$\pm$3~Myr; \citealp{2015arXiv150800898B}) known at the time.
(\citealp{rizzo14} classify the optical spectrum as M6$\beta$.) Cruz
\etal\ estimated a rough distance of 100~pc, and Gaia EDR3 finds it
resides at 148$\pm$5~pc \citep{2016A&A...595A...1G,
  2021A&A...649A...1G}.\footnote{The discrepancy arises because the
  source is significantly brighter than (old) field objects of similar
  spectral type, \eg, its \Ks-band absolute magnitude of
  7.60$\pm$0.08~mag is $\approx$1.6~mag brighter than field M6~dwarfs
  but similar to M6~\vlg\ objects
  \citep{2016ApJ...833...96L}.}\textsuperscript{,}\footnote{Cruz \etal\
  noted that another low-gravity late-M dwarf in their sample,
  2MASS~J0608528$-$275358, is close on the sky to \target\ but with a
  discrepant distance estimate of 30~pc. EDR3 places this object at
  44.2$\pm$0.3~pc, so indeed the two sources are not physically
  associated.}

\citet{2013ApJ...772...79A} classified the near-IR spectrum of \target\
as M5. Though this subclass was outside the early-type boundary of their
quantitative gravity classification scheme, they noted that the spectrum
qualitatively showed low-gravity features in comparison to field M5
dwarfs. Their low-resolution ($R\sim150$) spectrum, obtained on UT 2008
November~28, showed an unusually red near-IR continuum suggestive of
significant circumstellar reddening. Allers \& Liu estimated an
extinction of $A_V = 6.5$~mag from comparison with young objects of
similar spectral type. As the \citet{1998ApJ...500..525S} dust map
suggests very little extinction along this line of sight
($A_V \lesssim 0.2$~mag), such extinction is expected to be
circumstellar, consistent with the object's very-red near-IR continuum
and the finding from \citet{2012AAS...21934527L} of the object having a
mid-IR excess.

\citet{2016ApJ...833...96L} obtained a second low-resolution near-IR
spectrum on UT 2015 December~24, with the goal of higher S/N compared to
the 2008 spectrum from \citet{2013ApJ...772...79A}. Liu \etal\
classified their new spectrum as M6~\vlg. Surprisingly, the two spectra
were noticeably different, with the 2015 data being significantly bluer
than the 2008 data. \citet{2018ApJ...858...41Z} analyzed the two near-IR
spectra using their system for determining both spectral type and
reddening. They found spectral types of M5.3$\pm$0.9 and M5.8$\pm$0.9
and extinctions of $A_V = 6.3\pm4.0$ and $5.5\pm4.0$~mag for the 2008
and 2015 spectra, respectively. (The $\pm$4~mag uncertainty in $A_V$
adopted by the Zhang \etal\ methodology is conservative, based on the
intrinsic scatter in the colors for ultracool dwarfs. The two spectra
clearly have different continua.)

This paper is a follow-up study of \target, which verifies its near-IR
spectroscopic variability and demonstrates its photometric variability
at optical and mid-IR wavelengths. Section~2 presents new multi-epoch
spectra and photometry. Section~3 discusses the properties of the new
observations, and Section~4 concludes with a discussion of the system's
properties and comparison with other known objects.

%----------------------------------------------------------------------%
\section{Observations}

\subsection{Spectroscopy: IRTF/SpeX}

\begin{figure}
  \includegraphics[width=5in,angle=90]{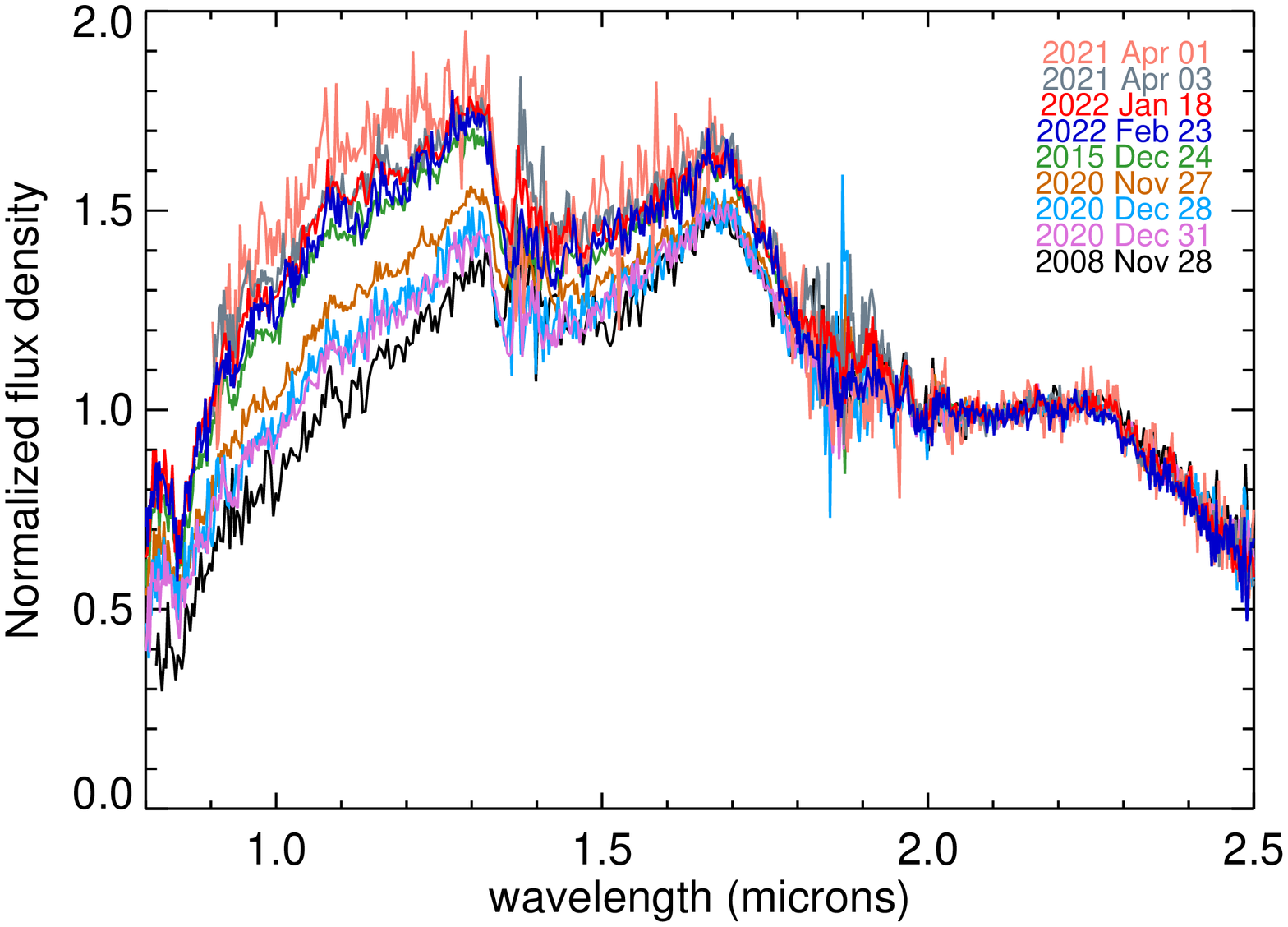}
  \caption{\normalsize IRTF/SpeX prism spectra of \target, all
    normalized to their 2.05--2.25~\micron\ flux. The legend
    with the observing dates is ordered by synthesized $J-K$ color, with the bluest
    epoch listed at the top. \label{fig:spex}}
\end{figure}

We obtained more near-IR \new{(0.8--2.5~\micron)} spectroscopy for
\target\ using the NASA Infrared Telescope Facility (IRTF) on the summit
of Mauna Kea, Hawaii. We obtained data on 5 separate nights spanning
about 4 months from November 2020 to early April 2021 \new{and then an
  additional 2 epochs in Jan/Feb
  2022}. We used the facility spectrograph SpeX
\citep{2003PASP..115..362R} in prism mode with the 0.5\arcsec\ slit,
which provided an average spectral resolution
($R\equiv\Delta\lambda/\lambda$) of $\approx$150. We dithered in an ABBA
pattern on the science target to enable sky subtraction during the
reduction process. For each epoch, we obtained calibration frames (arcs
and flats) and observed an A0~V star contemporaneously for telluric
calibration. All spectra were reduced using the Spextool software
package \citep{2003PASP..115..389V,2004PASP..116..362C}. \new{We
  determined near-IR spectral types and \citet{2013ApJ...772...79A}
  gravity classifications using the reddening-independent modification
  to the Allers \& Liu system developed by \citet{2018ApJ...858...41Z}.
  The spectral types between the different epochs are consistent with
  each other
  given the typing uncertainties. Most of the spectra have \vlg\ gravity
  classification, while the two \intg\ epochs have spectra
  with S/N below the $\gtrsim$50--75 needed for robust gravity
  classification, so there is no clear evidence for gravity class
  variability.} Table~\ref{table:spex} summarizes our observations and
derived measurements. Figure~\ref{fig:spex} plots our new spectra and
the published spectra, totalling 9~epochs spanning about 12.5~years.

\subsection{Photometry from Wide-Field Sky Surveys} \label{sec:photometry}

We examined several multi-epoch wide-field surveys to search for
photometric variability of \target. Photometry from the Pan-STARRS\,1
(PS1) telescope was extracted from observations obtained in two periods.
Observations prior to 2014~April~1 were performed as part of the
Pan-STARRS\,1 Science Consortium 3$\pi$~Survey
\citep{2016arXiv161205560C}. After 2014~April~1, observations were
performed as part of its NASA-funded mission to search for near-Earth
asteroids. Exposures from both periods were photometrically calibrated
with respect to the Pan-STARRS internal photometry database as described
by \cite{2020ApJS..251....6M}. Exposures from these two periods that
overlapped the field of interest were ingested into a stand-alone
photometry database following the process outlined in
\cite{2020ApJS..251....3M}. All measurements associated with the target
were extracted from the database. Of 227 measurements available, 5 were
rejected due to PSF\_QF $<$ 0.85 and 2 were rejected on the basis of bad
photometry flag bits SIZE\_SKIPPED and MOMENTS\_FAILURE
\citep{2020ApJS..251....5M}.
For epochs where \target\ was not detected by the standard processing,
the mask images were examined. The PSF-weighted mask fraction
(equivalent to PSF\_QF) was measured for the position of the target from
each exposure. All values fell into two ranges, $<0.05$ and $>0.95$. For
the former, the source position was masked, and thus the source could
not have been detected. For the latter, then \target\ should have been
detected if it were bright enough. The $5\sigma$ detection limit was
calculated as described in \cite{2020ApJS..251....5M}, with artificial
sources injected into the image near the expected detection threshold
and recovered if possible by the detection algorithm.
Figure~\ref{fig:ps1} shows the PS1 light curves spanning 2010
February~16 to 2021 February~22, giving both detections and 5$\sigma$
upper limits. Variability is seen on a wide range of timescales, from
intra-night to over the entire 11-year dataset.

The ongoing All-Sky Automated Survey for SuperNovae (ASAS-SN;
\citealp{2014ApJ...788...48S, 2017PASP..129j4502K}) and the Asteroid
Terrestrial-impact Last Alert System (ATLAS;
\citealp{2018PASP..130f4505T}) provide wide-field data with less
sensitivity than PS1 but much higher cadence. Neither of these surveys
is sensitive enough to detect \target's photosphere, but we examined
their datastream to look for bright flaring.
ASAS-SN observed \target\ for a total of 956 epochs in $V$~band (358
epochs) or $g$~band (598 epochs) from UT 2013 November~1 through UT 2021
December~6. The median $V$- and $g$-band 5$\sigma$ upper limits are 17.9
and 18.4~mag, respectively, without any detections.

ATLAS observed \target\ for a total of 98~epochs in the $c$~band
(420--650~nm) and 308 epochs in the $o$~band (560-820~nm) between UT
2015 October~22 and UT 2021 December~3 at a typical cadence of 4~days.
There are no ATLAS detections, with median 5$\sigma$ upper limits of
20.2 in the $c$~band and 19.6 mag in the $o$~band.

\new{No photometric variability is reported in \Gaia\ DR2 for \target\
  from the {\tt phot\_variable\_flag} field. \citet{2021ApJ...912..125G}
  have developed a method to detect variable sources from the \Gaia\
  photometry catalogs based solely on the reported $G$-band fluxes, flux
  errors, and number of observing epochs. Computing their {\sc varindex}
  statistic (see their Appendix~E) for DR2 and EDR3 data gives values of
  0.075 and 0.048, respectively. For comparison, they find values of
  $>$0.007 and $>$0.0027 for the top 1\% of the most variable sources in
  their white dwarf sample. The values for \target\ far exceed these
  criteria, suggesting \target\ is variable in $G$~band.}

The AllWISE variability flag is {\tt 11nn}, indicating no significant
evidence for mid-IR variability over the two epochs of observations in 2010--2011.
\new{CatWISE2020 \citep{2021ApJS..253....8M} provides 14 epochs of data
  taken from 2010--2018 and reports p-values for $W1$ and $W2$ of 0.027
  and \ex{9.8}{-7}, respectively, suggesting significant mid-IR
  variability, especially in the longer wavelength band. We examine the
  individual epochs in Section~\ref{sec:phot-midIR}.}

\subsection{LCOGT Optical Photometric Monitoring}

The \target\ field was imaged through the Sloan $gri$ and Pan-STARRS $z$
passbands with the Spectral camera on the Faulkes-South 2-m telescope of
the Las Cumbres Global Telescope network
\citep[LCOGT,][]{2013PASP..125.1031B} at 30~epochs between UT 2021
December~29 and 2022 February~15. The Spectral camera provides a
10.5\arcmin\ $\times$ 10.5\arcmin\ field of view with 0.152\arcsec\
pixels binned to 2$\times$2. Integration times were 200, 35, 40, and 160
sec in the respective $griz$ bandpasses to achieve a photon-noise
S/N=200 in all but the $g$ band, for which we expect S/N$\approx$20 for
\target. Bias removal, flat-fielding, source identification, and
aperture photometry were performed by the automated {\tt BANZAI}
pipeline \citep{2018SPIE10707E..0KM}. The faintness of \target\ in
$g$-band meant that it was usually not detected by {\tt BANZAI}, and
thus aperture photometry was performed directly at the location of the
source. Relative time-series photometry was calculated by matrix
solution of a set of linear equations describing the relation between
instrumental magnitudes and apparent (top of the atmosphere) magnitudes
with observation zero-points, a second-order (color-dependent)
extinction coefficient, and an instrument-specific color-term. (We used
Gaia EDR3 $G_{BP}-G_{RP}$ colors for the latter two terms.) Reference stars
are selected based on brightness, color, and lack of significant
variability as estimated by \Gaia, and excluding \target\ itself. The
residuals of each individual reference star are calculated after
solution of the matrix relation; the reference stars with the largest
residuals (after subtraction of formal errors) are removed from the fit;
and the fit re-calculated. This iteration continues until the median
error in the zeropoints, which dominates the errors in the final
photometry, is minimized.

Figure \ref{fig:lcogt} shows the relative photometry in each of the four
passbands. Variability between the passbands is not consistent,
suggesting that the origin is not astrophysical, except perhaps for the
case of the Sloan~$r$ filter. This bandpass contains the H$\alpha$ line,
which is likely variable given the star's youth. The overall variability
(median absolute deviation) in the $griz$ passbands is 0.30, 0.11,
0.045, and 0.10 mag, respectively, but this is not much different than
the median formal errors (photon noise and zero-point error added in
quadrature) of 0.22, 0.08, 0.038, and 0.036~mag, respectively. The
corresponding reduced $\chi^2$ values for each filter's photometry
relative to the mean values are 4.0, 4.0, 1.7, and 14.6, suggesting
general constancy in flux except for $z$-band (which could be affected
by fringing and thereby leading to a high $\rchisq$ value). We suspect
that $g$-, $r$-, and $z$-band uncertainties are actually somewhat
underestimated due to data systematics and that the star was not
significantly ($\gtrsim$0.1~mag) variable during the LCOGT observations.

\subsection{NTT/SOFI Near-IR Photometric Monitoring}

We obtained near-continuous IR photometry of \target\ on two consecutive
half-nights with the Son of ISAAC (SOFI) instrument, the IR spectrograph
and imaging camera on the 3.58-m ESO New Technology Telescope (NTT). We
observed \target\ in \Js\ band on the night of 2022 February~19 UT with
a total elapsed time of 3.9 hours and interleaved in \Js, $H$ and \Ks\
bands on the following night of 2022 February~20 UT with a total elapsed
time of 4.2 hours. The \Js\ band has a central wavelength of
1.240~\micron\ and a width of 0.290~\micron; the $H$ band is at
1.653~$\mu$m with a width of 0.297~$\mu$m; the \Ks\ band is at
2.162~$\mu$m with a width of 0.275~$\mu$m. The field of view of SOFI is
4.92\arcmin\ $\times$ 4.92\arcmin\ with a pixel scale of
0.288\arcsec/pixel. We used an ABBA nod pattern during the observations
in order to subtract the background, with three 60-second images taken
at each nod position. The observations were obtained entirely at
airmass less than 2. The seeing was 0.8--1.3\arcsec\ on the first
night and 0.6--1.2\arcsec\ on the second night.

We followed the same data reduction as \citet{2019MNRAS.483..480V}. We
corrected the inter-quadrant row crosstalk and applied flat-fielding and
illumination corrections. Frames of one nod were subtracted with the
frame of the opposite nod closest in time to remove the background. Then
we applied aperture photometry to the stars detected in the field. We
selected a fixed aperture radius that was similar to the median FWHM of
all stars and also that minimized the photometric noise of the target
light curve. The photometric noise was measured by the same method used
in \citet{2014ApJ...793...75R}, namely the standard deviation of the
light curve subtracted by a shifted light curve of itself with one time
stamp difference, divided by $\sqrt{2}$. The final aperture radius we
used for the 2022-02-19 \Js-band data was 1.0\arcsec\ and was
1.15\arcsec\ for the 2022-02-20 \Js-, $H$- and \Ks-band data. The
light curve extracted at each nod position was normalized by its median.
We used the iterative algorithm of \citet{2014ApJ...793...75R} to select
a set of reference stars that were not variable after being detrended by
other reference stars. Then we took the median light curve of these
reference stars as the calibration light curve. The light curve of the
target was divided by this calibration light curve in order to remove
the systematic variability caused by the instrument and the atmosphere.
Figure~\ref{fig:sofi} shows the final detrended light curves. Some
low-amplitude ($\approx$1\%) variability appears to be present over the
$\approx$4~hour observations, especially in the \Js-band light curve
from the first night and perhaps the \Ks-band light curve of the second
night, and is plausibly due to photospheric inhomogeneities, \eg, a
hot/cool spot, instead of indicating hour-scale extinction variations.

We analyzed the detrended light curve for each filter using the
Lomb-Scargle periodogram \citep[LS;][]{1976Ap&SS..39..447L,
  1982ApJ...263..835S} but did not detect a significant periodicity
within the observing duration. We also ran the LS periodogram on the
joined two-night light curve of \Js\ band, finding a peak power at 4.11
hours. To check false-positive detections caused by our observation
cadence, we also ran the LS periodogram on our observation window
function, which is the same time series as the light curve but with a
unit value of 1 \citep{2018ApJS..236...16V}.
Though the 4.11~hour period did not correspond a strong peak in the
window function's periodogram, the joined \Js-band periodogram also has
peaks of similar power at other similar periods, so we do not consider
this to be a definitive periodicity detection. A longer-duration
continuous dataset would be valuable.

%----------------------------------------------------------------------%
\section{Analysis}

\subsection{Spectroscopy}

\subsubsection{Variable Near-IR Extinction \label{sec:nearIR}}

Visual examination of the near-IR spectra suggests that the observed
changes in the continuum slope could be due to extinction changes. To
quantify this, we measured the relative extinction between different
epochs, using the highest S/N spectrum (from December 2015) as a
template. For each of the other epochs, we fitted the template to the
data assuming the template had been reddened by $A_V$ that followed the
\citet{1999PASP..111...63F} extinction law with the usual
$R = A_V / E(B-V) = 3.1$, \new{along with a wavelength-independent
  scaling factor to account for differences in the overall flux
  calibration due to varying slit losses and source brightness compared
  to the 2015 template.} We performed the fit using a grid of $A_V$
and scale factors, adopting the values corresponding to the lowest
\chisq. (We also tried fits with $R$ as an additional free parameter.
The resulting constraints were poor, as expected given the limited
wavelength range of our spectra.) The formal uncertainties on $A_V$ from
the fits were negligible, but conservatively assuming an uncertainty for
SpeX spectra's continua of $\Delta(J-K)=0.05$~mag
\citep[e.g.][]{2012ApJS..201...19D,
  2012ApJ...750..105R} would imply an uncertainty of
$\sigma(A_V)=0.3$~mag. For our four epochs from 2021 and 2022, the fits
resulted in a negative value for $A_V$, as expected given that these
epochs' spectra are bluer than the 2015 template. Table~\ref{table:spex}
includes the fitting results, and Figure~\ref{fig:dered} shows the
dereddened spectra. The latter indicates that the differences between
the near-IR spectra are consistent with changes in their relative
extinction by $\Delta{A_V}\approx$2~mag.

\new{Note that the timescale for extinction changes can be quite
  heterogenous. For
  instance, the two observations from 2020 December have very similar
  inferred extinctions while the two from 2021 April have notably
  different ones, even though each pair of observations is separate by
  only a few days. In contrast, our final 3 epochs of spectra, spanning
  10.5 months, have nearly consistent spectra, suggesting the system may
  have been in a steady state for nearly the past year.}

\subsubsection{Optical Spectrum}

Comparison of the \citet{2003AJ....126.2421C} optical spectrum to the
solar-metallicity M6 template from PyHammer~v2.0
\citep{2017ApJS..230...16K} supports the notion that \target\ has
significant reddening. Figure~\ref{fig:dered-optical} shows that a
reddening of $A_V=4.3$~mag is needed to make the spectra agree, an
extinction roughly in accord with the published $A_V\approx6$~mag
estimates.\footnote{\new{In principle, we should also compare to
  unreddened young M6 objects to gauge the extinction, but there is a
  paucity of suitable data. The RIZzo spectral library \citep{rizzo14}
  only contains two young objects with M6 spectral types besides
  \target.
  2MASS~0557--1359 (optical spectral type M6:$\gamma$ in RIZzo; M7 in
  \citealp{2007AJ....133..439C-others}) is a somewhat enigmatic young
  field object with a near-IR spectrum that shows lower surface gravity
  (\ie, younger age) than most field objects and a very bright absolute
  magnitude \citep{2016ApJ...833...96L}, making it a poor comparison
  object.
  2MASS~2234+4041 (optical spectral type M6:$\delta$) is a very young
  ($\sim$1~Myr) member of the LkH$\alpha$233 star-forming group
  \citep{2009ApJ...697..824A}, so its optical spectrum is likely to
  differ from a young field M6 dwarf due to both extinction and gravity
  effects.
  \citet{2018ApJS..234....1B} show that the optical PS1 colors for young
  late-M dwarfs are similar to those of field late-M dwarfs, so using
  the PyHammer~v2.0 template is a reasonable choice for estimating the extinction of
  \target.}}

Using the \citet{2003AJ....126.2421C} optical spectrum, we measure an
equivalent width (EW) of $-7.4\pm0.4$~\AA\ for the H$\alpha$ emission,
from simple Gaussian fitting assuming a constant continuum level and
Monte Carlo error propagation.
\new{For more direct comparison with the literature, we also measure
  EW(H$\alpha$) of $-8.9\pm2.7$~\AA\ following the definitions of the
  emission line and its continuum level from
  \citet{2015AJ....149..158S} (with the lower S/N of this measurement
  compared to direct Gaussian fitting being due to the narrow continuum
  bands.)}
For spectral types M5 and M6, \citet{2003AJ....126.2997B} set 18.0 and
24.1~\AA, respectively, as the H$\alpha$ equivalent width that
distinguishes objects with strong chromospheric activity from those with
accretion, based on low-resolution optical spectra for members of
star-forming regions and young ($\lesssim$125~Myr) clusters. Thus
\target\ is not clearly accreting. \new{Also, the measured EW(H$\alpha$)
  for \target\ is not particular distinctive for age estimation, as
  mid-M dwarfs with ages spanning $\sim10^{7-9}$~yr can show such
  values (\eg, Figure~5 of \citealp{2021AJ....161..277K}).}

%------------------------------------------------------------%
\subsection{Photometry}
%------------------------------------------------------------%

\subsubsection{Optical Variability}

In contrast to our NTT/SOFI data, the PS1 light curves suggest possible
short-timescale ($\approx$1-hour) variability.
Figure~\ref{fig:shorttime} shows the time blocks with 4 or more
photometry points separated by $<$1~hour, taken in the same filter, and
having significant non-constant flux (specifically, having p-values
$<10^{-4}$, indicating the block's data are not consistent with a
constant flux given the measurement uncertainties). Such rapid changes
could be due to intrinsic photospheric variability (cool/hot spots or
localized accretion features; \eg, \citealp{1994AJ....108.1906H,
  2011ApJ...731...17B}) or extrinsic variability due to small-scale
inhomogeneties in the circumstellar material. \new{We caution that
  some deviations may be due to instrumental systematics, \eg,
  when the object falls on two different detectors on the same night, as
  the PS1 \TPS\ observing strategy was not optimized for high-precision
  short-term photometric monitoring.} High-cadence, multi-wavelength
monitoring would help constrain the possible scenarios.

\begin{figure}
  \vskip 0.5in
  \includegraphics[width=5in,angle=90]{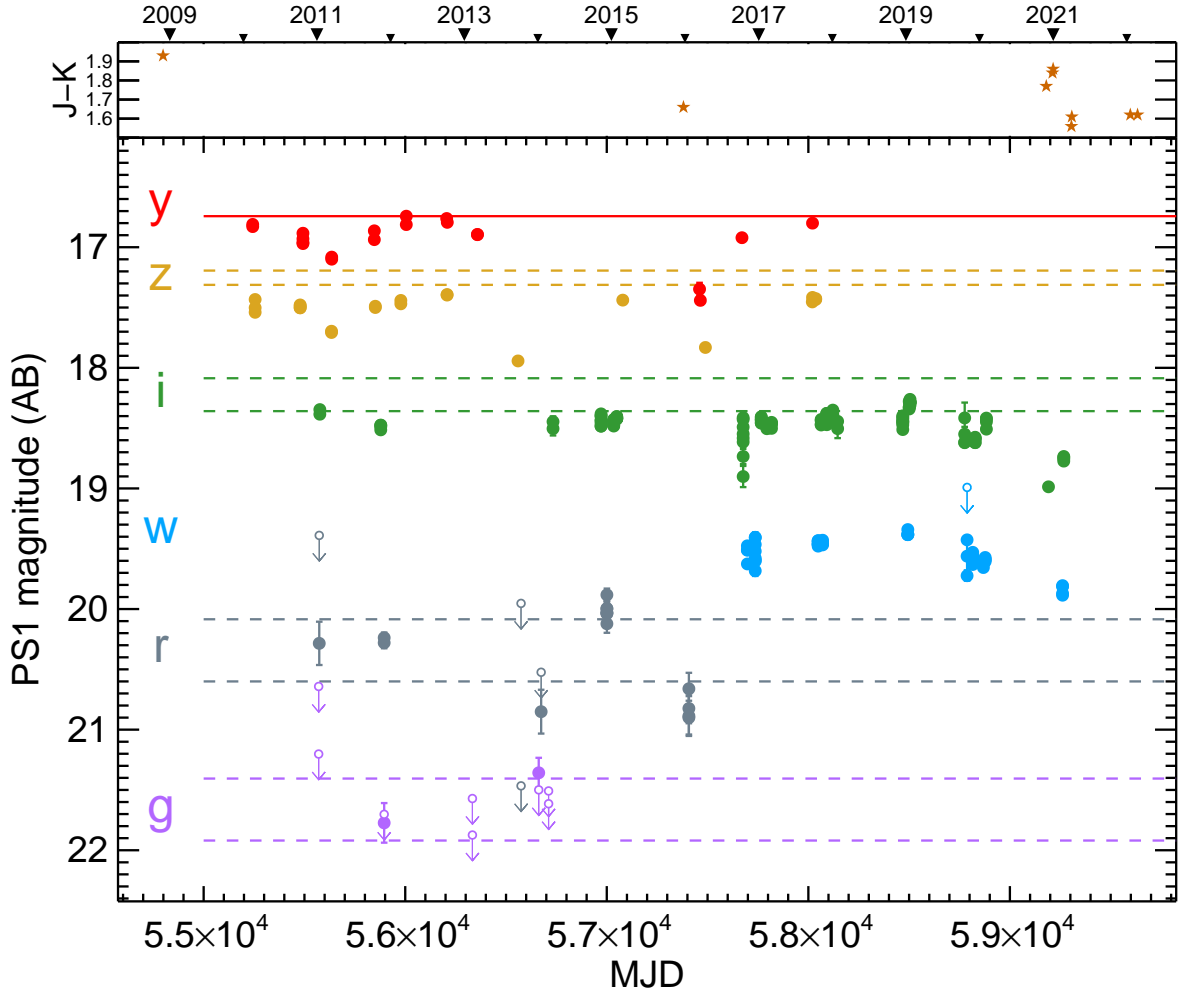}
  \caption{\normalsize PS1 light curve for \target, showing variabilty
    on timescales from sub-day to 11~years. \new{Upper limits
      (5$\sigma$) are plotted as open circles with downward-pointing
      arrows. For reference, the dashed lines indicate the predicted
      1$\sigma$ range of magnitudes assuming no extinction, adopting the
      PS1 colors of M6 dwarfs from \citet{2018ApJS..234....1B}, and
      using the minimum (brightest) \yps\ magnitude (which is shown as
      the solid red line) to compute magnitudes from the colors. Such
      color information is not available for the \wps\ band, so there
      are no corresponding dashed lines for it. The small panel at the
      top shows the $(J-K)_{MKO}$ color synthesized from our SpeX
      spectra, and labels at the very top denote the start of each
      calendar year.} \label{fig:ps1}}
\end{figure}

\begin{figure}
  \centering{\includegraphics[width=6in,angle=0]{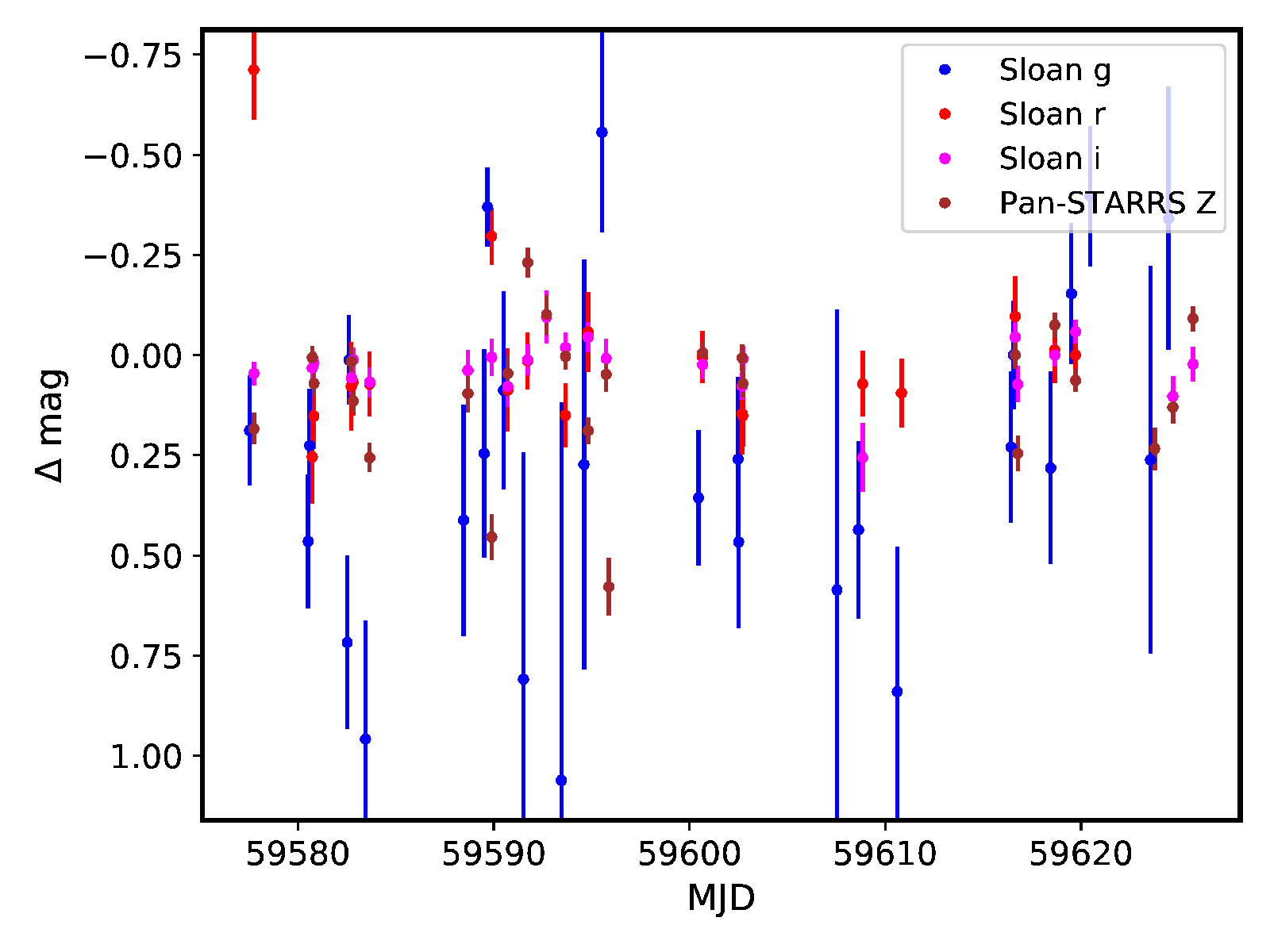}}
  \caption{\normalsize \new{Optical light curves of \target\ taken with
      LCOGT, showing relative photometry of the source from December
      2021 to February 2022. For each filter, the baseline flux is taken
      to be the median of the upper 50\% of points and subtracted from
      all the epochs. The different pass-bands offset by 0.2 days for
      clarity. No clear evidence for variability is detected over this
      1.5-month duration given the measurement uncertainties.}
    \label{fig:lcogt}}
\end{figure}

\begin{figure}
  \vskip 1in
  \centering{\includegraphics[width=6.5in,angle=0]{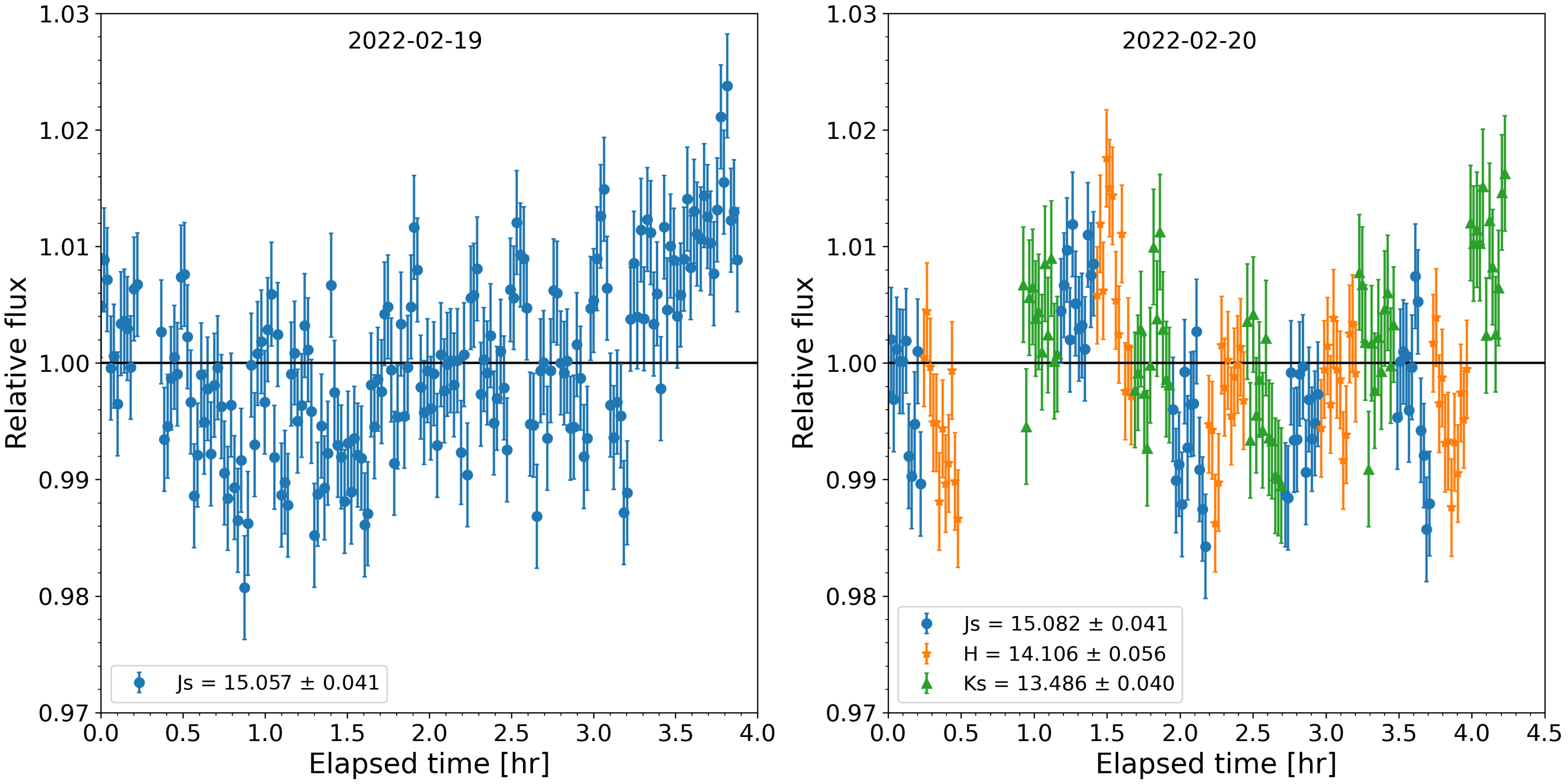}}
  \vskip 2ex
  \caption{\normalsize \new{Near-IR light curves of \target\ taken with SOFI in \Js,
    $H$ and \Ks\ bands. The light curves show variability with an
    amplitude of $\approx$1\% over the $\approx$4 hours of observing at
    each night. The median magnitude of each light curve is listed in
    the legend, as calibrated with 2MASS stars in the field.}
  \label{fig:sofi}} \end{figure}

\begin{figure}
  \includegraphics[width=5in,angle=90]{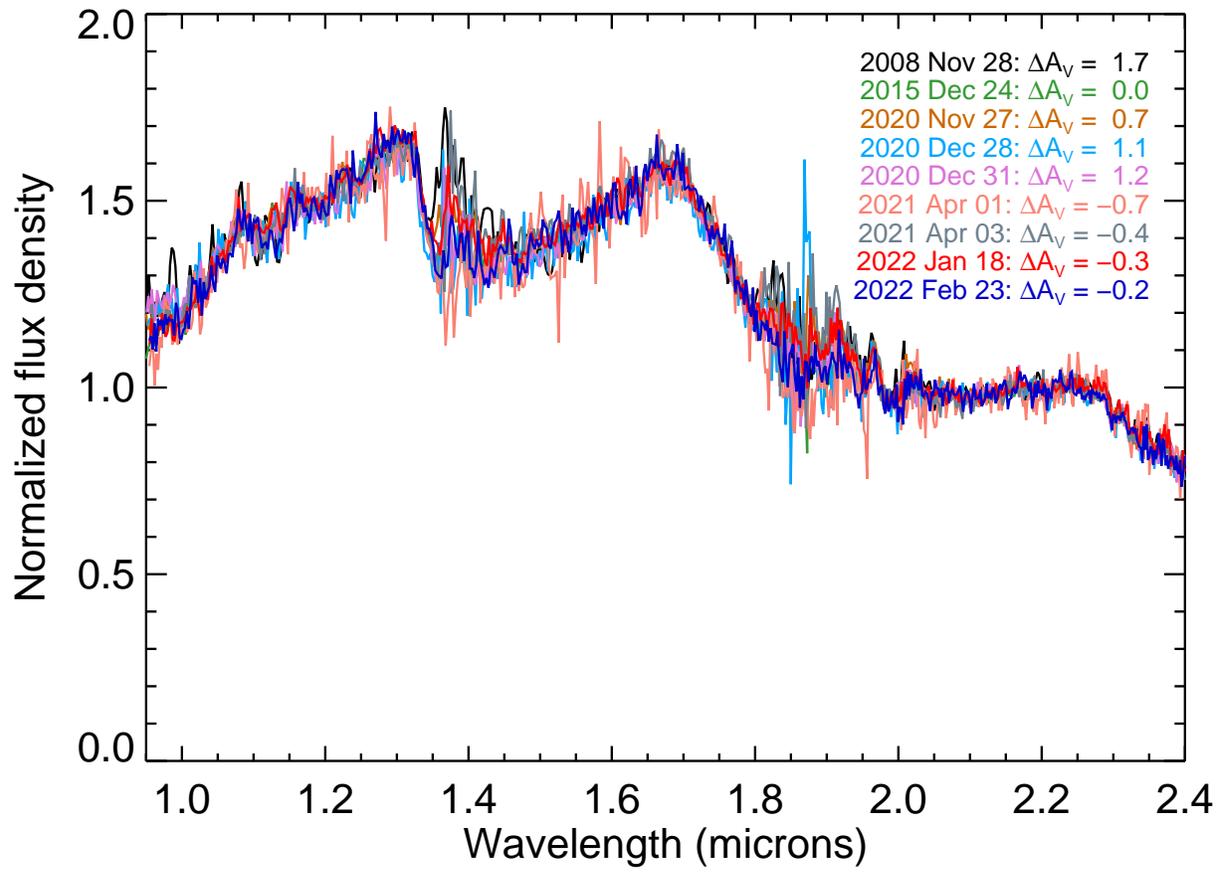}
  \caption{\normalsize IRTF/SpeX prism spectra of \target\ after
    dereddening with the best-fitted extinction relative to the 2015
    spectrum, all normalized to their 2.15--2.25~\micron\ 
    flux. The results are consistent with the scenario that the near-IR
    variability is due to changes in circumstellar
    extinction. \label{fig:dered}}
\end{figure}

\begin{figure}
  \includegraphics[width=5in,angle=90]{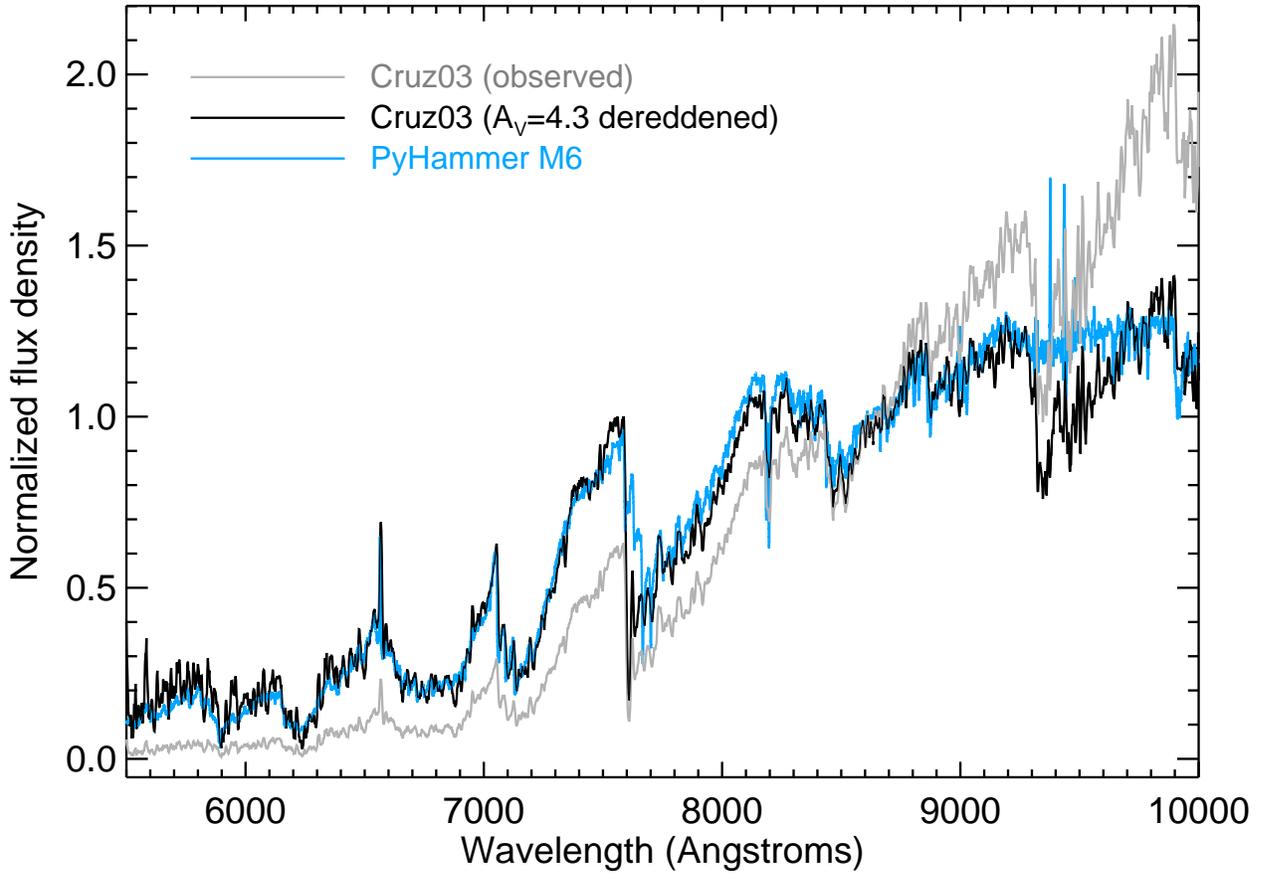}
  \caption{\normalsize Comparison of \target\ optical spectrum from
    \citet{2003AJ....126.2421C} with the solar-metallicity M6 template
    from PyHammer v2.0 \citep{2020ApJS..249...34R}, normalized by their
    fluxes at 8400--8900~\AA. The observed spectrum ({\em grey}) is
    significantly redder than the template. After dereddening by
    $A_V=4.3$~mag, the agreement is much better. (The Cruz \etal\ spectra
    plotted here have been smoothed by a 5-pixel boxcar.) \label{fig:dered-optical}}
\end{figure}

\begin{figure}
  \centering{\includegraphics[width=4.7in,angle=90]{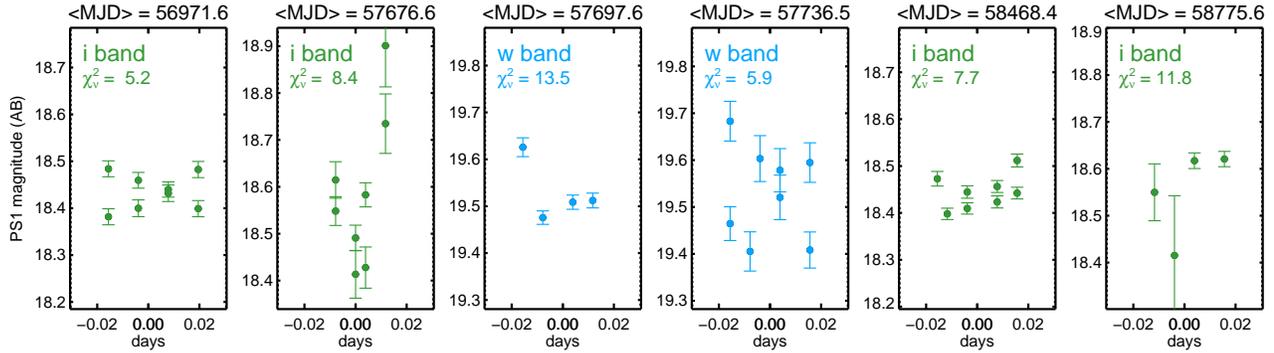}}
  \vskip -1.3in
  \caption{\normalsize Possible short-timescale variability of \target\
    from PS1 optical photometry, \new{showing the 6 most variable
      observing blocks}
    (defined as 4 or more detections in the same filter within a 1.5-hour window).
    Different colors indicate different filters, and the titles give the
    average Modified Julian 
    Date for each block's detections. The reduced \chisq\ value for the
    photometry compared to each block's average magnitude is also
    reported, all of which correspond to p-values of $<10^{-4}$. The
    $x$- and $y$-axes ranges are identical in all the   
    plots. \label{fig:shorttime}}
\end{figure}

Figure~\ref{fig:ps1} shows the PS1 \zps- and \yps-band photometry
changes are seemingly correlated over the many years of data, namely
both become faint or bright at the same time. The non-simultaneity of
the observations precludes a more definitive statement, as photometry in
the two filters were typically taken 1--2~weeks apart during the
\TPS.\footnote{There
  are only 2 blocks of PS1 photometry for which multi-filter data were
  obtained $\le$2 days apart, both involving the \zps\ and \yps\
  filters. The two epochs give colors of \zps--\yps of 0.51$\pm$0.02~mag
  ($<$MJD$>$ = 55634.8, $\Delta{t}$=1.1 days) and 0.62$\pm$0.02~mag
  ($<$MJD$>$ = 56297.4, $\Delta{t}$=2.0 days).} Over all 11~years of
data, the full ranges of the \zps\ and \yps\ magnitudes are 0.6 and
0.7~mag, respectively, corresponding to $A_V$ changes of 1.1 and 1.8~mag
using the \citet{2012ApJ...750...99T} extinction curve.\footnote{The
  Tonry \etal\ extinction curve for PS1 depends on the \gps--\ips\
  color, though the dependence is very weak for the standard bandpasses.
  Using the typical $\gps-\ips$ color for M5--M6.9 dwarfs from
  \citet{2018ApJS..234....1B}, the extinction values ($A_\lambda$) in
  \grizy\ are factors of 1.10, 0.80,
  0.60, 0.48, and 0.40 different from that of $V$ band ($A_V$).}
\new{Similarly, the full ranges of \rps\ and \ips\ magnitudes are 1.0
  and
  0.7~mag,
  respectively, corresponding to $A_V$ changes of 0.8 and 0.4~mag
  (though \rps\ could be affected by variable H$\alpha$ emission).} Such
extinction variations are comparable in amplitude to those inferred from
the near-IR spectra (Table~\ref{table:spex}).

\new{In addition to variability, the PS1 light curves suggest there
  might be periods of very low/no extinction.} The mean \zps\ and \yps\
magnitudes from the \TPS\ were 17.524$\pm$0.006 and
16.933$\pm$0.017~mag, respectively, giving a color of
$\zps-\yps=0.59\pm0.02$~mag, similar to the median color of
0.44--0.51~mag for M5.0--M6.9 dwarfs from \citet{2018ApJS..234....1B}.
For the same spectral type range, Best \etal\ tabulate median colors of
$\rps-\ips=1.9-2.1$~mag and $\ips-\zps=0.9-1.0$~mag, comparable to the
colors of $\rps-\ips=1.71\pm0.05$~mag and $\ips-\zps=0.967\pm0.008$~mag
computed from the non-simultaneous \TPS\ photometry, \ie, no strong
reddening. (Note there are relatively few epochs of \ips\ data from the
\TPS, as many of the pointings resulted in the source landing on a
masked portion of the detector.) \new{To further illustrate this,
  Figure~\ref{fig:ps1}
  shows
  the
  predicted magnitudes of \target\ assuming the brighest \yps-band
  detection represents the unextincted photosphere and using the
  \citet{2018ApJS..234....1B} colors for M6~dwarfs. The brightest
  detections in the other bands are just consistent with the predicted
  unextincted magnitudes, suggesting some epochs have low extinction.}

The interpretation of the bluest (\gps-band) detections is unclear.
There are two modest ($S/N~=~6$ and~8) detections of \target, both of
which are verified by visual inspection of the reduced images. Both of
these images were obtained as one of a pair of images taken 17~minutes
apart, given the standard PS1 3$\pi$~Survey observing strategy of
searching for fast-moving asteroids by taking pairs of closely-spaced
exposures. For the detection on MJD 55895.49280 of 21.77$\pm$0.16~mag,
the immediately preceding image had a 5$\sigma$ detection limit of
\gps=21.70~mag, and visual inspection of the smoothed image shows no
sign of any counterpart. For the detection on MJD 56662.38760 of
21.36$\pm$0.12~mag, \target\ was not detected to a depth of
\gps=21.5~mag in the other image of the pair, which we corroborate with
visual inspection. This suggests that $g$-band variability of
$\gtrsim$0.2~mag occurred on $\approx$20~minute timescales, consistent
with flaring activity of field M~dwarfs \citep[e.g.][]{
  2014ApJ...797..122D, 2016ApJ...829...23D}.

\new{However, given the large inferred extinction for the source, the
photosphere should not have been detected at all in \gps~band.
The \citet{2018ApJS..234....1B} compilation of M~dwarfs shows that M6
dwarfs have $\gps-\Ks$ colors of 7.2$\pm$0.3~mag.
Assuming $A_V=6$~mag ($A_{\gps}\approx7$, $A_K=0.7$~mag), the 2MASS
\Ks-band photometry of \target\ would imply $\gps\approx20\pm0.3$~mag
and $\approx$27~mag for its unextincted and extincted photosphere,
respectively. Thus if the large optical extinctions inferred by
\citet{2013ApJ...772...79A} and \citet{2018ApJ...858...41Z} from the
near-IR spectra are typical, then \target\ is expected to be undetected
by PS1 in \gps~band, given the 5$\sigma$ limiting magnitudes of 22.0 and
23.3~mag for the PS1 single-epoch and stacked exposures, respectively.
Thus the two \gps-band detections can be attributed to flaring activity.
The implied flare amplitudes would be $\Delta\gps\gtrsim4$~mag,
consistent with the largest flares seen for field M~dwarfs
\citep{2019ApJ...876..115S}. On the other hand, the unextincted
photosphere is well above the \gps-band limiting magnitudes, and thus
the two $\gps\approx21.5$~mag detections could arise from epochs when
the photosphere was only lightly extincted. More intensive multi-band
monitoring could distinguish between flaring and extinction variations.}

\subsubsection{Mid-IR Variability}  \label{sec:phot-midIR}

\begin{figure}
  \centering
  \includegraphics[width=3.5in,angle=90]{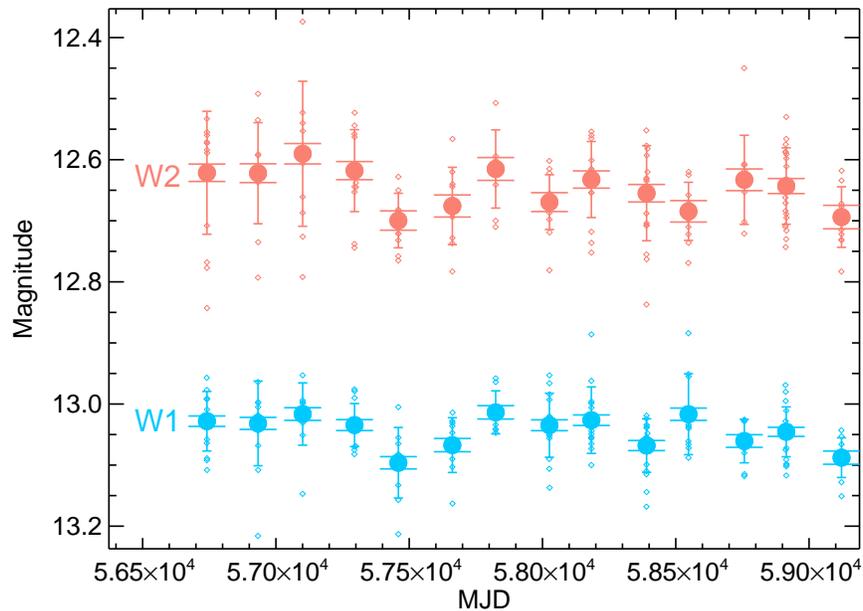}
  \caption{\normalsize \NEOWISE\ mid-IR light curve of \target\
    from March 2014 to September 2020 using single-exposure photometry, with the weighted
    average at each visit plotted as a large circle with short, wide-hatted error bars. The
    tall, small-hatted error bars show the RMS photometric scatter at each visit. The intra-visit scatter on
    day-long timescales is consistent with constant flux given the measurement uncertainties,
    while the inter-visit scatter on 6-month timescales suggests possible $\approx$0.1~mag variability
    (Section~\ref{sec:phot-midIR}). 
    \label{fig:neowise}}
\end{figure}

We investigated the mid-IR varability of \target\ indicated by
CatWISE2020 (Section~\ref{sec:photometry}) by examining the individual
single-exposure photometry reported by the \NEOWISE\ Reactivation
Mission \citep{2014ApJ...792...30M, https://doi.org/10.26131/irsa144},
spanning 2014 March~24 to 2020 September~30. The object is well detected
in individual exposures, with $S/N$ of 34 and 20 in $W1$ and $W2$,
respectively. Following the recommendations in
\citet{2015nwis.rept....1C}, we removed any data with non-zero confusion
flags ({\tt cc\_flags}), low quality scores ({\tt qual\_frame}=0 or
{\tt qi\_fact}=0), small distance from the South Atlantic Anomaly ({\tt
  saa\_sep}$<$5), or close to the moon ({\tt moon\_masked}$\ne$0). We
also removed epochs with poor reduced \chisq\ values for the profile-fit
photometry ($\rchisq>6$), though these were relatively few after
applying the aforementioned criteria. Figure~\ref{fig:neowise} plots the
\NEOWISE\ light curves.

These data probe two distinct timescales, each day-long visit having
$\approx$10--20~exposures and then the visits being spaced 6~months
apart. There is not much evidence for significant intra-visit (day-long)
variability. Compared to a model of constant flux at each visit, 3
visits of $W1$ data and 2 visits of $W2$ data out of the 14 total visits
have p-values $<$\ex{1}{-3}. In contrast, there is potential evidence
for variability between the 6-month visits, with p-values of
$<$\ex{1}{-6} for both $W1$ and $W2$ when comparing the individual
visits' weighted averages to a model of constant flux during all
14~visits.

\begin{figure}
  \hbox{
    \hskip -0.3in
    \includegraphics[width=3.5in,angle=90]{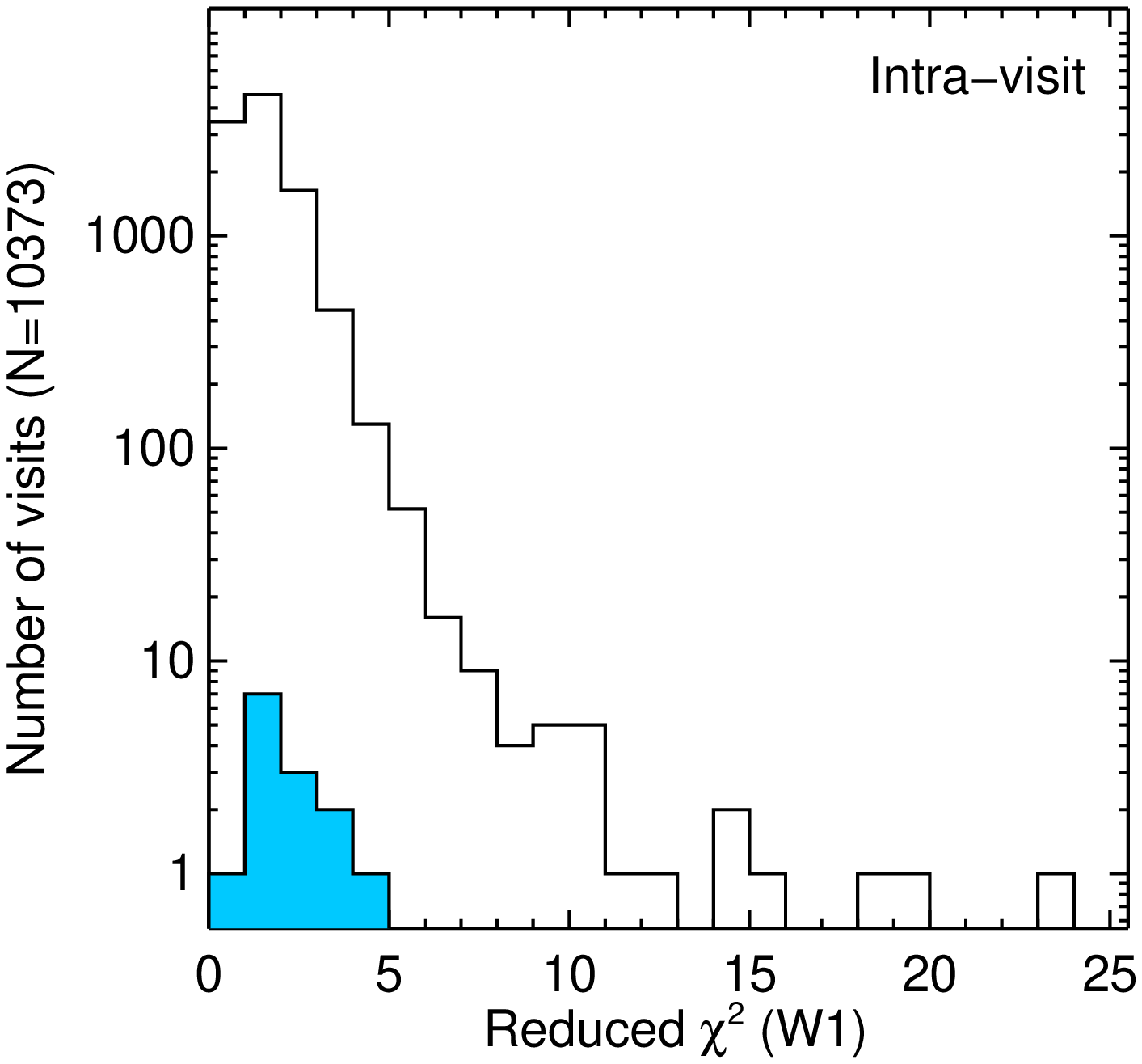}
    \hskip -1.7in
    \includegraphics[width=3.5in,angle=90]{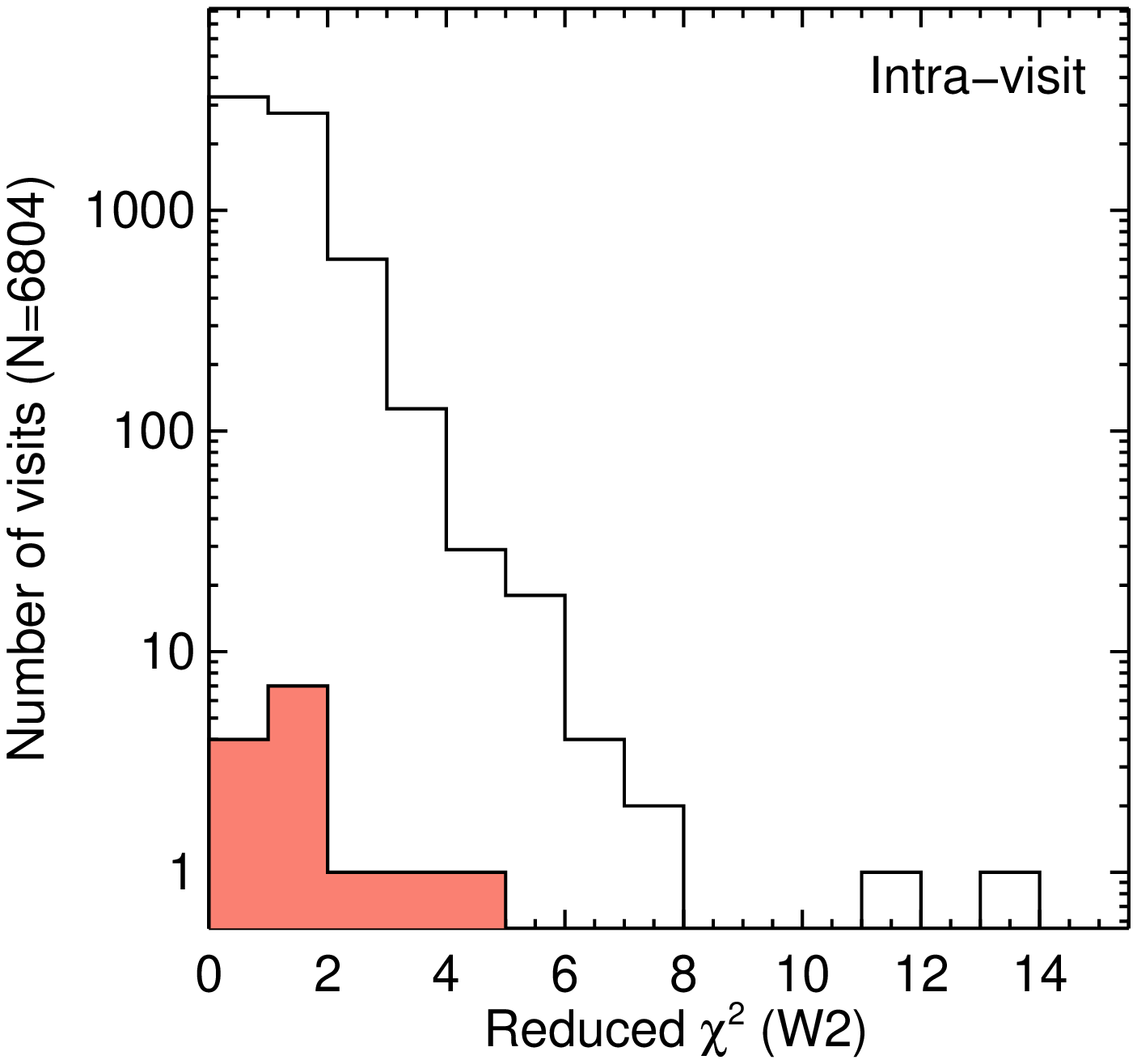}
  }
  \hbox{
    \hskip -0.3in
    \includegraphics[width=3.5in,angle=90]{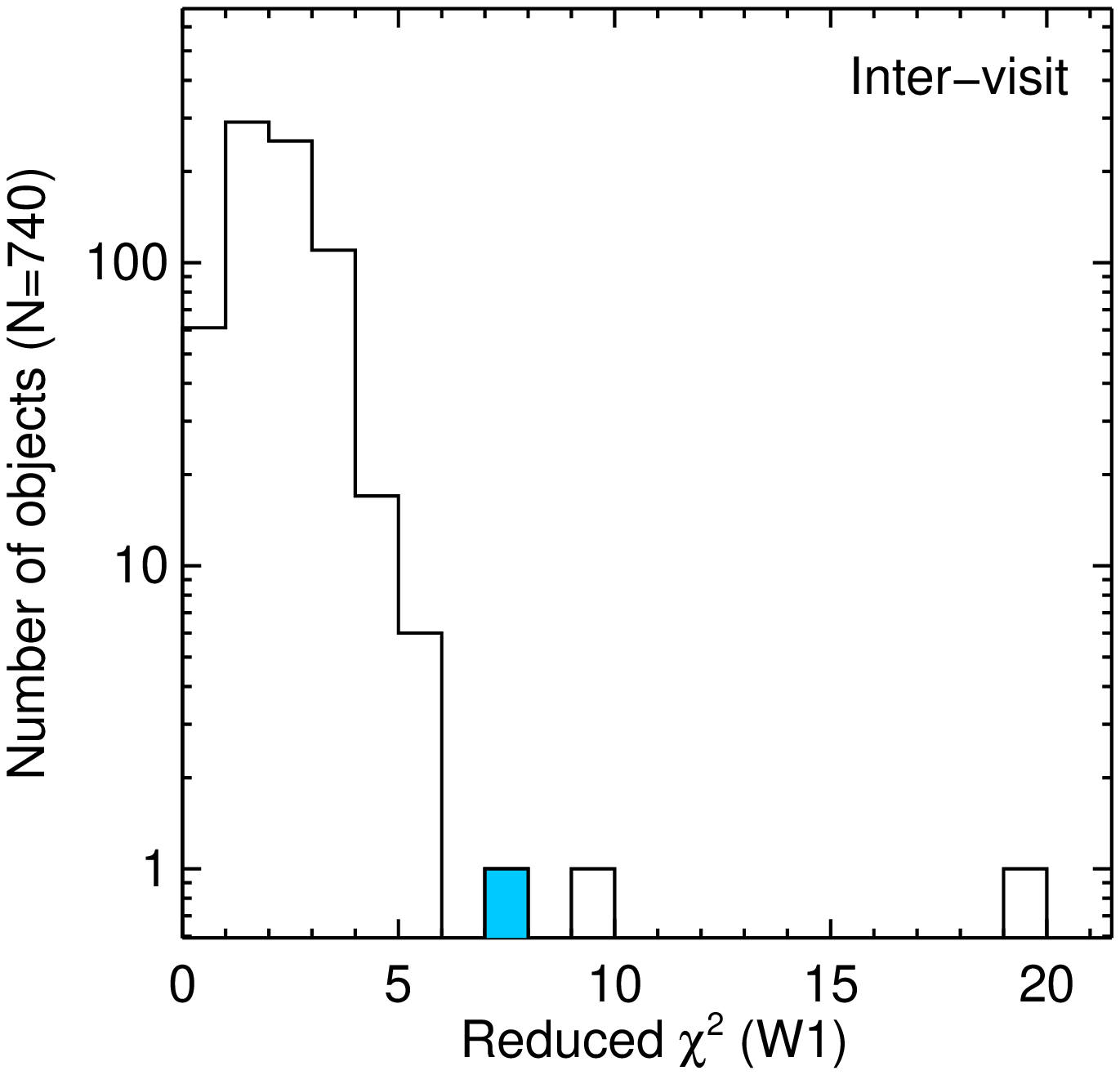}
    \hskip -1.7in
    \includegraphics[width=3.5in,angle=90]{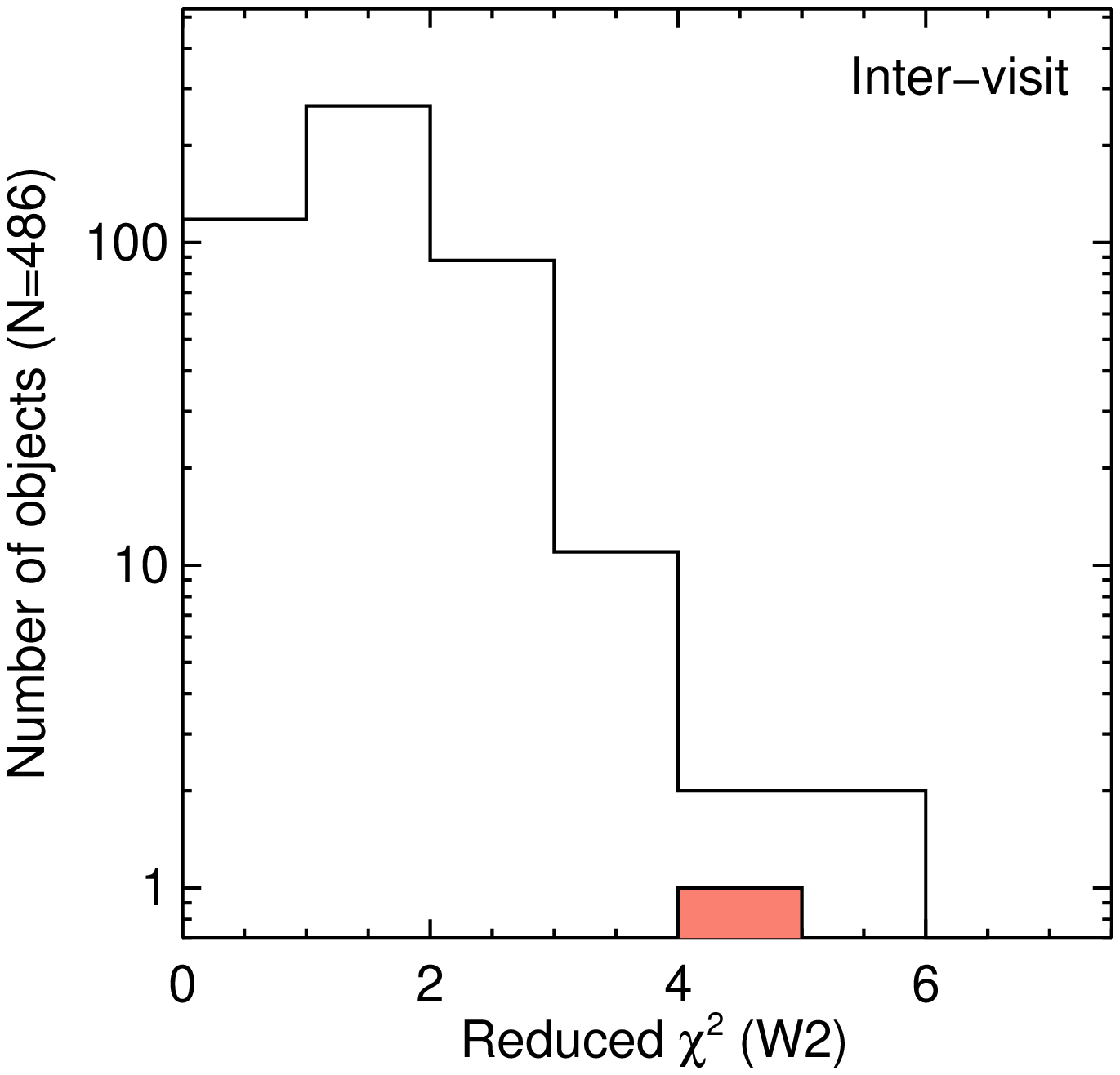}
  }
  \caption{\normalsize Comparison of \target's mid-IR
    variability compared to sources of comparable magnitude
    ($\pm$0.25~mag) and within 1~deg. {\em Top:} Reduced \chisq\ values
    for each day-long visit compared to a model of constant flux. (Each
    object has $\approx$1~dozen visits.) The histogram shows that
    \target's variations are undistinguished, \ie, no significant
    evidence for variability. {\em Bottom:} Reduced \chisq\ values for
    each object compared to a model where the weighted averages at each
    visit are consistent with constant flux. \target\ has the 3rd and
    4th highest \rchisq\ values, suggesting possible mid-IR variability.
    (For all plots, note that the number of exposures and/or visits is
    not the same for every object, so these histograms cannot be directly
    compared to a specific \chisq\ distribution.)
    \label{fig:neowise-chisq}}
\end{figure}

To investigate further, we retrieved the \NEOWISE\ single exposures for
all sources within 1\degs\ of \target\ and having similar ($\pm$0.25~mag)
photometry in either $W1$ or $W2$. For each bandpass, after applying the
same data-quality cuts, we computed the same \rchisq\ statistics for the
intra-visit and inter-visit fluxes as we did for \target.
Figure~\ref{fig:neowise-chisq} shows the resulting distributions. For
intra-visit data, \target\ seems unremarkable relative to the comparison
sample. In contrast, its inter-visit photometry suggests that it is
variable on $\sim$6-month timescales, as its $W1$ and $W2$ \rchisq\
values are the 3rd and 4th highest out of all the sources in the field.

Taking the mid-IR variability at face value, the $\approx$0.1~mag
changes would correspond to extinction variations of
$A_V\approx2.5-4.0$, comparable to the values inferred from our
optical/NIR data. Alternatively, the mid-IR variations could also arise
from changes in the disk emission, likely from the inner wall location
at the dust sublimation radius \citep[e.g.][]{2013AJ....145...66F}.
Simultaneous optical/NIR photometry during \NEOWISE's future visit of
\target\ would be valuable to distinguish between variations in
extinction and emission.

% - - - - - - - - - - - - - - - %
\subsection{Astrometry, Age, and Mass}

\citet{2014ApJ...783..121G} found that the proper motion of \target\
suggested possible membership in the Columba moving group. But a
preliminary (S/N~$\approx$~2) parallax from \citet{2016ApJ...833...96L}
was inconsistent with the inferred distance of 55~pc if such membership
were correct. The BANYAN~$\Sigma$ \citep{2018ApJ...856...23G} algorithm
finds that this object's Gaia EDR3 astrometry indicates that it is a
field object, not belonging to a known young stellar group or
association.

We queried \Gaia~EDR3 \citep{2016A&A...595A...1G,2021A&A...649A...1G} to
search for possible wide companions to \target. Only two stars within
2\degree (5.1\,pc) of \target\ have a similar parallax ($\pm1$\,mas) and
proper motion ($\pm1$\,mas\,yr$^{-1}$). The closer of the two (Gaia~EDR3
2898355055033597184, PSO~J094.9535$-$29.0808) is only 1\farcm2 away and
has a parallax within 0.14~mas (0.6$\sigma$) and proper motion differing
by 0.36\,mas\,yr$^{-1}$ (1.6$\sigma$). This star was identified by
\citet{2021ApJ...917...23K} as being young ($31^{+22}_{-10}$\,Myr), but
not a member of any known or newly identified young associations. This
star and \target\ have also been noted as a highly likely physical pair
by \citet{2021MNRAS.506.2269E} with a chance-alignment parameter
$<0.01$. Such pairs have an average chance alignment probability of
0.08\%, and the youth of both objects makes them even more unlikely to
be a chance alignment. We thus consider \target\ to be a physical
companion to this star and adopt the star's $\approx$8$\times$ more
precise parallax and proper motion for the whole system. This places
\target\ at a distance of $145.2\pm0.6$\,pc with a projected separation
of $10450\pm50$\,AU from its host star. We use the PARSEC~v1.2S
isochrones \citep{2015MNRAS.452.1068C} and the
\citet{2021ApJ...917...23K} age, assuming negligible extinction based on
the 3D map of \citet{2017A&A...606A..65C}, to estimate a mass of
$0.30^{+0.04}_{-0.03}$\,\Msun\ for the primary star using its absolute
magnitude of $M_G=9.970\pm0.009$\,mag.

The other star found in our 2\degree-radius query (Gaia~EDR3
2898345984062275072) is 1\fdg521 away and differs significantly in
proper motion ($0.89\pm0.03$\,mas\,yr$^{-1}$) from \target. At such a
large separation (3.8\,pc), projection effects of space motion into
astrometric measurements are substantial, and indeed we can find a good
match with the $UVW$ space motion of both this wide object and the
\target\ system if we assume a radial velocity of 25--30\,km/s for all
three stars. However, this star is not identified as young by
\citet{2021ApJ...917...23K}, even though it would be a
$0.44\pm0.04$\Msun\ pre-main sequence star if it were the same age as
the \target\ system. We thus consider this wider star to be a chance
alignment.

The comoving companion of \target\ provides an age for the system,
enabling a mass estimate using evolutionary models. We use the 2MASS
\Ks-band magnitude and a bolometric correction for young objects from
\citet{2015ApJ...810..158F} to estimate the bolometric luminosity of
\target. Adopting a uniform age distribution spanning 21--37~Myr, the
\citet{2015A&A...577A..42B} models give a mass of $0.11\pm0.02$~\Msun\
for \target, assuming no extinction and using a Monte Carlo approach to
account for the uncertainties in the magnitude, bolometric correction,
distance, and age. If we assume an extinction of $A_V=6$~mag, the
resulting mass is $0.17\pm0.03$~\Msun.

%---------------------------------------------------------------------%
\section{Discussion \label{sec:discussion}}

\new{The variability exhibited by \target\ is not typical for field
  late-M dwarfs. Such stars display photometric modulations due to
  rotation of their spotted photospheres, showing semi-amplitudes of
  $\approx$0.2--2.0\% and periods of $\approx$1--100~days, with
  kinematically older stars tending to have longer periods than
  kinematically younger stars \citep{2016ApJ...821...93N,
    2018AJ....156..217N}. This differs from the much larger-amplitude
  ($\approx0.5$~mag) optical variability and the near-IR extinction
  variations shown by \target, with the variability demonstrating
  timescales spanning days/weeks (IRTF/SpeX and PS1), months (PS1 and
  \NEOWISE), and years (PS1). (Our one continuous $\approx$4-hour dataset
  from SOFI does show low-amplitude variability consistent with a
  spot-modulated rotation light curve.)}

\new{Around young ($\sim$30-100~Myr) field stars comparable in age to
  \target, a small fraction ($\sim$1\%) of stars with debris disks show
  extremely large IR excesses suggestive of transient collisional events
  \citep{2009ApJ...698.1989B, 2012Natur.487...74M, 2021ApJ...910...27M}.
  Some of these show large-amplitude ($\approx$30--100\%) mid-IR
  variability with systematic trends on $\sim$year timescales but
  without any changes in their optical photometry
  \citep{2015ApJ...805...77M, 2021ApJ...918...71R}. This is unlike
  \target, where both optical and mid-IR variability occurs, the
  variability has a larger amplitude in the optical than the mid-IR, and
  the mid-IR variability has a $\sim$0.1~mag amplitude without any trend
  -- all these suggest that it does not possess a transient debris
  disk.}

\new{Stars in young ($\approx$1--15~Myr) star-forming regions show a
  much richer set of variability phenomena, caused by both photospheric
  and circumstellar sources, which more resemble the behavior of
  \target.
  In addition to rotation-induced modulations from cool and hot spots,
  photometry of young stars exhibits accretion-related fluctuations,
  variable obscuration/color changes, and quasi-periodic/aperiodic
  dimming (``dipping'') events with timescales of less than a few days,
  indicating origins at the photosphere or inner disk
  \citep[e.g.,][]{1994AJ....108.1906H, 2001AJ....121.3160C,
  2014AJ....147...83S, 2016ApJ...816...69A}.
On timescales of $\sim$years, optical photometry for most young stars is
relatively stable, with only a small fraction exhibiting long-term
changes due to variable extinction \citep{2007A&A...461..183G,
  2008A&A...479..827G, 2017MNRAS.465.3889R, 2019MNRAS.483.1642H}. In
contrast, long-timescale variability at mid-IR wavelengths is more
common ($\sim$30--50\%; \eg, \citealp{2012ApJ...748...71F,
  2014AJ....148...92R, 2021ApJ...920..132P}) and may arise from
structure fluctuations in the inner disk, variable accretion, and/or
variable extinction.
Finally, a few young systems have very complex light curves suggestive
of significant occultation by a structured circumstellar or
circumplanetary disk, \eg, AA~Tau \citep{2013A&A...557A..77B}, V409 Tau
\citep{2015AJ....150...32R}, KH~15D \citep{2001ApJ...554L.201H,
  2006ApJ...644..510W}, 1SWASP~J140747.93-394542.6
\citep{2012AJ....143...72M, 2014MNRAS.441.2845V} and PDS~110
\citep{2017MNRAS.471..740O}. Continuous monitoring of \target\ is needed
to determine whether its extinction variations could arise from such
exotic configurations, which cannot be discerned from our existing
sparse-cadence data.}

\begin{figure}
  \hskip -0.75in
  \hbox{
    \includegraphics[width=3.75in,angle=90]{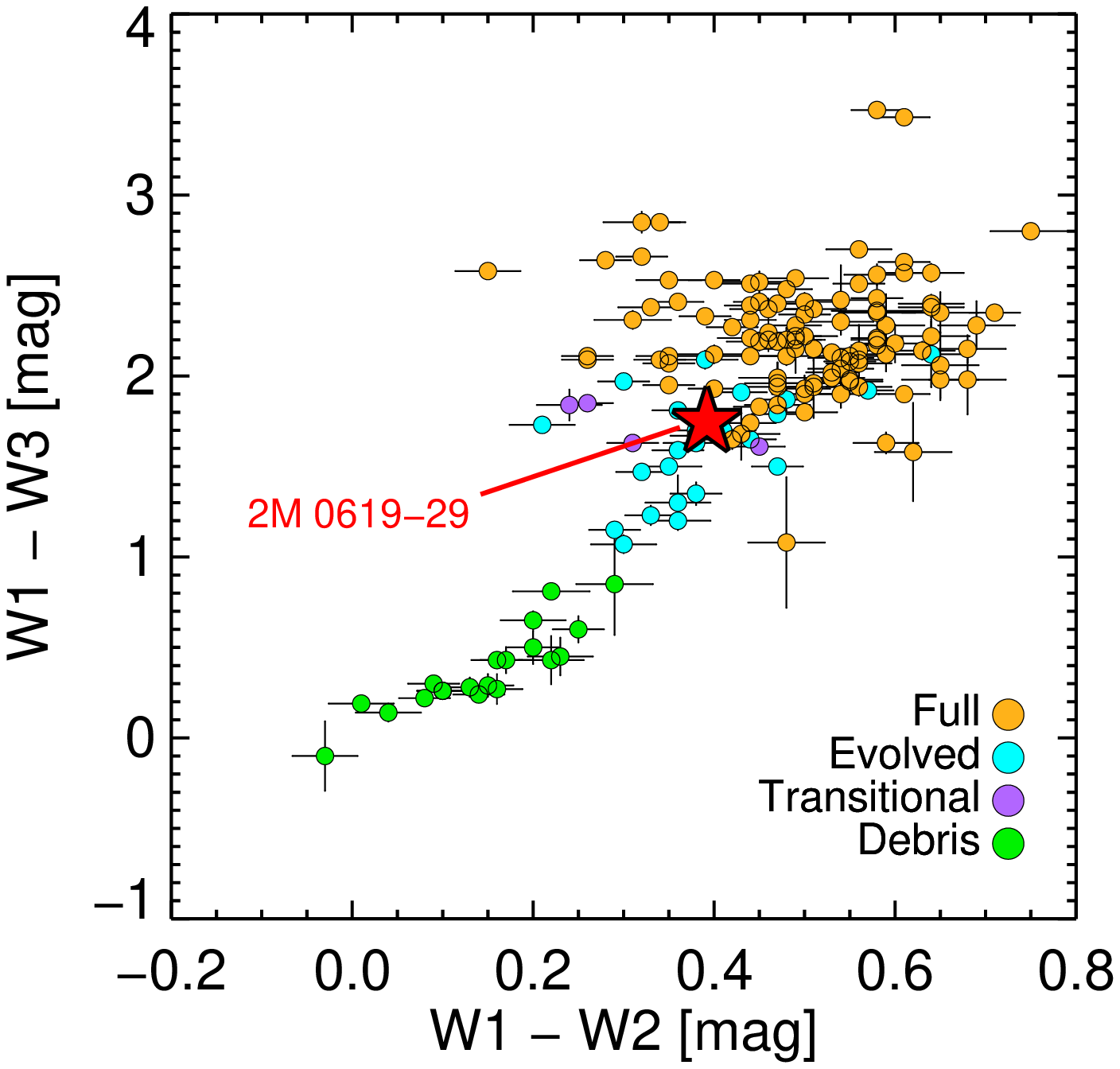}
    \hskip -1.75in
    \includegraphics[width=3.75in,angle=90]{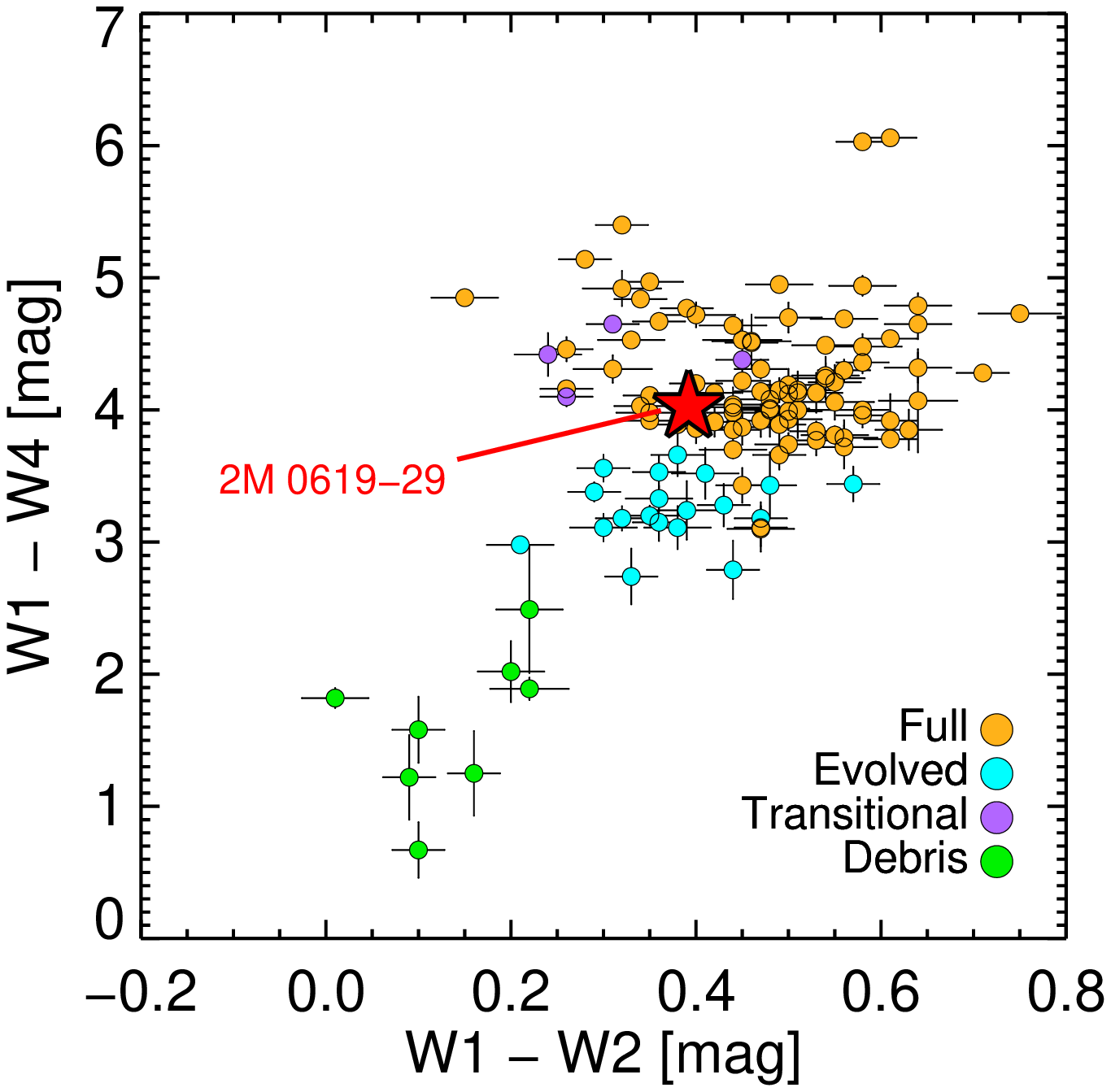}
  }
  \caption{\normalsize \new{Comparison of the AllWISE mid-IR photometry for
    \target\ with disk-bearing M~dwarfs in the Upper Sco region
    ($\approx$11~Myr) from \citet{2012ApJ...758...31L}.  The colors
    represent different evolutionary states of the circumstellar disks.
    The colors show that \target\ possesses either a full or evolved disk,
    which is unexpected given its $\approx$30~Myr age. \label{fig:disk}}}
\end{figure}

The overall properties of \target's circumstellar material suggests a
disk in between the primordial disks abundant in young
($\lesssim$10~Myr) star-forming regions and the debris disks seen around
older stars. The inferred $A_V$ values are too large to be associated
with a debris disk, where even edge-on systems present little
obscuration of the central star, while the H$\alpha$ emission is not
strong enough to clearly show ongoing accretion. The AllWISE colors of
\target\ ($W1-W2=0.43\pm0.03$, $W1-W3=1.75\pm0.12$ and
$W1-W4=4.04\pm0.4$~mag) suggest emission from \new{an evolved or full
  circumstellar disk, instead of a debris disk (Figure~\ref{fig:disk})}.
(From the nomenclature of, \eg, \citet{2012ApJ...747..103E}, evolved
disks are gas-rich primordial disks in the process of becoming optically
thin but which do not possess large gaps or holes like transitional
disks.) Note that the wavelength gap between $W2$- and $W3$-band
photometry means an inner disk gap might not be noticed with the
existing photometry.

Substantial disks older than $\sim$20~Myr are rare \new{(\eg,
  \citealp{2014AJ....147..146K} find a frequency of $<$0.8\% for
  optically thick and thin disks around M~dwarfs in the $\sim$40~Myr old
  Tuc-Hor Association)}, but they do exist. \new{In addition to \target,
  \citet{2013ApJ...772...79A} and \citet{2016ApJ...833...96L} identified
  3 M~dwarfs that straddle the stellar/substellar mass boundary and have
  strong IR excesses, though these may be very young ($\sim$Myr) sources
  based on their absolute magnitudes.}
V4046~Sgr is a classical T~Tauri K5+K7 binary with a gas-rich evolved
circumbinary disk \citep{1997AJ....114..301J, 2008A&A...492..469K,
  2010ApJ...720.1684R}, belonging to the $\beta$~Pictoris moving
group ($\sim$25~Myr).
\citet{2009ApJ...702L.119R} identified a young, accreting brown dwarf
that they suggested to be in the $\sim$40~Myr Tuc-Hor Association (and
reaffirmed by \citealp{2014ApJ...783..121G}); we re-confirmed this
object's Tuc-Hor membership (99.6\% probability) using \Gaia\ EDR3
astrometry and the radial velocity from \citet{2017AJ....154...69S} as
inputs to the BANYAN~$\Sigma$ algorithm \citep{2018ApJ...856...23G}.
\citet{2013ApJ...774..101R} found two young ($\sim10-100$~Myr) field
M~dwarfs with strong mid-IR excesses ($W1-W4\approx4$~mag, comparable to
\target).\footnote{\new{\citet{2013ApJ...774..101R} find possible
    Columba ($\sim$40~Myr) membership for these two sources, 2MASS
    J02590146$-$4232204 and 2MASS J03244056$-$3904227, based on their
    kinematics and estimated distances. Using the latest astrometry from
    Gaia EDR3 as input to BANYAN $\Sigma$, we find the former is a field
    object (96.4\% probability) and the latter is a Tuc-Hor member
    (99.7\%; also identified as a member by
    \citealp{2014AJ....147..146K}).}}
\citet{2016ApJ...832...50B} identified two young brown dwarfs in the
Columba and Tuc-Hor moving groups ($\sim$40~Myr) with strong accretion
and large fractional disk luminosities, and also highlight the Argus
($\sim$40~Myr) membership of the T~Tauri K5+K7 binary FK~Ser
\citep{1996ApJ...458..312J}.
\citet{2016ApJ...830L..28S} identified an M~dwarf member of the Carina
($\sim$45~Myr) moving group with a strong \WISE\
excess.\footnote{\citet{2018MNRAS.476.3290M} find this object,
  WISE~J080822.18-644357.3, shows significant mid-IR variability and
  possesses strong broad H$\alpha$ emission indicating ongoing
  accretion. Its \WISE\ colors suggest a (gas-rich) primordial disk as
  opposed to a (gas-poor) debris disk, but ALMA observations by
  \citet{2019ApJ...872...92F} establish the disk is gas-poor and has a
  relatively low disk mass compared to primordial disks. They suggest
  its circumstellar dust is generated from a collisional cascade of
  km-sized bodies in the outer ($<$16~AU) disk and then migrates inward
  due to Poynting-Robertson drag to produce the strong mid-infrared
  excess.}
\new{\citet{2020MNRAS.494...62L} identified an M~dwarf member of the Argus
($\sim$55~Myr) moving group with a strong \WISE\ excess and ongoing
accretion.}
\new{\citet{2022arXiv220414163G} identified another M~dwarf member of
  the Carina moving group with a long-lived primordial disk showing
  ongoing accretion and with a \TESS\ light curve having many dipping
  events. They also suggest such long-lived M~dwarf disks could be the
  progenitors of multi-planet systems around low-mass stars akin to
  TRAPPIST-1 \citep{2017Natur.542..456G}.} 

\citet{2020ApJ...890..106S} brand these long-lived systems as ``Peter
Pan'' disks, namely unusually long-lived ($\gtrsim20$~Myr) accretion
disks around M~dwarfs with strong infrared excesses ($K-W4>2$~mag).
\citet{2020MNRAS.496L.111C} suggest that for primordial disks around
low-mass stars to be so long-lived, the disks must have possessed a
combination of rare intrinsic and/or extrinsic properties, such as low
disk viscosity, extremely low external photoevaporation (indicating
formation in rare environment), and relatively high disk masses.
\citet{2022MNRAS.509...44W} find that primordial disk lifetimes as long
as $\sim$50~Myr can only occur around $\lesssim$0.6~\Msun\ stars.
\target\ fulfills most of the \citeauthor{2020ApJ...890..106S}\ criteria
for these objects. Though it lacks a clear signature of accretion based
on its EW(H$\alpha$), some of the Peter Pan objects would also not be
classified as accretors if solely based on their EW(H$\alpha$), as their
accretion is only detected with higher resolution spectra that resolves
the H$\alpha$ linewidths. The long-term variability of the Peter Pan
sample is not yet known, and \target's behavior suggests such data could
be quite interesting. Overall, \target\ may be a more distant, more
extincted version of a Peter Pan disk.

More intensive monitoring of \target\ is needed to better understand the
geometrical, physical, and dynamical nature of its variabilty. Perhaps
most notably, our existing data compilation lacks any long-term
continuous monitoring (aside from the 1.5~months of LCOGT optical
monitoring, which shows little variability), and a more extensive
dataset could explore possible hour-scale variations suggested by the
PS1 and NTT photometry. Such a continuous dataset would also help to
constrain the orbital radius and spatial structure of the circumstellar
material, through periodicity, color, and extinction measurements.
Higher spectral resolution optical H$\alpha$ data would be a sensitive
probe for ongoing accretion. Finally, longer wavelength sub-mm/radio
observations would better assess the evolutionary state of \target's
disk by measuring the gas and dust masses.

%----------------------------------------------------------------------%
\begin{acknowledgments}

  Conducted amidst the turbulence of Fall 2020, the discussions in the
  ASTR~633 Astrophysical Techniques class at the University of Hawaii
  were the genesis of this study, resulting in a third epoch of spectra
  that demonstrated the variability of \target\ seen by between the 2008
  and 2015 spectra of \citet{2016ApJ...833...96L} continues to persist.
  We thank John Rayner, Bobby Bus, and Michael Connolly for supporting
  the IfA graduate students' IRTF proposal writing and observing time.
  This work has benefited from discussions with Rob Siverd and John
  Tonry.
  This work has benefited from The UltracoolSheet at
  \url{http://bit.ly/UltracoolSheet}, maintained by Will Best, Trent
  Dupuy, Michael Liu, Rob Siverd, and Zhoujian Zhang, and developed from
  compilations by \citet{2012ApJS..201...19D},
  \citet{2013Sci...341.1492D}, \citet{2016ApJ...833...96L},
  \citet{2018ApJS..234....1B}, and \citet{2020AJ....159..257B}.
  This research has benefitted from the Ultracool RIZzo Spectral
  Library, maintained by Jonathan Gagn{\'e} and Kelle Cruz.
  This research was funded in part by the Gordon and Betty Moore
  Foundation through grant GBMF8550 and by the NSF through grant
  AST-1518339, both awarded to M.\ Liu.
  B.~Shappee is supported by NSF grants AST-1908952, AST-1920392 and
  AST-1911074.
  E.~Gaidos acknowledges support by NASA grants 80NSSC19K0587
  (Astrophysics Data Analysis Program) and 80NSSC19K1705 (TESS Guest
  Observer Cycle~2). 

  This work has made use of data from the European Space Agency (ESA)
  mission {\it Gaia} (\url{https://www.cosmos.esa.int/gaia}), processed
  by the {\it Gaia} Data Processing and Analysis Consortium (DPAC,
  \url{https://www.cosmos.esa.int/web/gaia/dpac/consortium}). Funding
  for the DPAC has been provided by national institutions, in particular
  the institutions participating in the {\it Gaia} Multilateral
  Agreement.
  This publication makes use of data products from the Wide-field
  Infrared Survey Explorer \citep{2010AJ....140.1868W}, which is a joint
  project of the University of California, Los Angeles, and the Jet
  Propulsion Laboratory/California Institute of Technology, and NEOWISE
  \citep{2011ApJ...731...53M}, which is a project of the Jet Propulsion
  Laboratory/California Institute of Technology. WISE and NEOWISE are
  funded by the National Aeronautics and Space Administration.
  The Pan-STARRS\,1 Surveys (PS1) and the PS1 public science archive
  have been made possible through contributions by the Institute for
  Astronomy, the University of Hawaii, the Pan-STARRS Project Office,
  the Max-Planck Society and its participating institutes, the Max
  Planck Institute for Astronomy, Heidelberg and the Max Planck
  Institute for Extraterrestrial Physics, Garching, The Johns Hopkins
  University, Durham University, the University of Edinburgh, the
  Queen's University Belfast, the Harvard-Smithsonian Center for
  Astrophysics, the Las Cumbres Observatory Global Telescope Network
  Incorporated, the National Central University of Taiwan, the Space
  Telescope Science Institute, the National Aeronautics and Space
  Administration under Grant No. NNX08AR22G issued through the Planetary
  Science Division of the NASA Science Mission Directorate, the National
  Science Foundation Grant No. AST-1238877, the University of Maryland,
  Eotvos Lorand University (ELTE), the Los Alamos National Laboratory,
  and the Gordon and Betty Moore Foundation.
  Pizza.
  This work has made use of data from the Asteroid Terrestrial-impact
  Last Alert System (ATLAS) project. The Asteroid Terrestrial-impact
  Last Alert System (ATLAS) project is primarily funded to search for
  near-Earth asteroids through NASA grants NN12AR55G, 80NSSC18K0284, and
  80NSSC18K1575; byproducts of the NEO search include images and
  catalogs from the survey area. This work was partially funded by
  Kepler/K2 grant J1944/80NSSC19K0112 and HST GO-15889, and STFC grants
  ST/T000198/1 and ST/S006109/1. The ATLAS science products have been
  made possible through the contributions of the University of Hawaii
  Institute for Astronomy, the Queen’s University Belfast, the Space
  Telescope Science Institute, the South African Astronomical
  Observatory, and The Millennium Institute of Astrophysics (MAS),
  Chile.
  % 
  % Funding for the SDSS and SDSS-II has been provided by the Alfred P.
  % Sloan Foundation, the participating institutions, the National
  % Science Foundation (NSF), the U.S. Department of Energy, the
  % National Aeronautics and Space Administration, the Japanese
  % Monbukagakusho, the Max Planck Society, and the Higher Education
  % Funding Council for England.
  %
  We thank Las Cumbres Observatory and its staff for their continued
  support of ASAS-SN. ASAS-SN is funded in part by the Gordon and Betty
  Moore Foundation through grants GBMF5490 and GBMF10501 to the Ohio
  State University, and also funded in part by the Alfred P. Sloan
  Foundation grant G-2021-14192. Development of ASAS-SN has been
  supported by NSF grant AST-0908816, the Mt. Cuba Astronomical
  Foundation, the Center for Cosmology and AstroParticle Physics at the
  Ohio State University, the Chinese Academy of Sciences South America
  Center for Astronomy (CAS- SACA), the Villum Foundation, and George
  Skestos.
  This work makes use of observations from the Las Cumbres Observatory
  global telescope network with Key Project 2020-007
  ``Catch A Fading Star.''
  Burrito.
  Our research has employed the 2MASS data products; NASA's
  Astrophysical Data System; and the SIMBAD database operated at CDS,
  Strasbourg, France.
  This research has made use of the NASA/IPAC Infrared Science Archive,
  which is funded by the National Aeronautics and Space Administration
  and operated by the California Institute of Technology.

  Finally, the authors wish to recognize and acknowledge the very
  significant cultural role and reverence that the summit of Mauna Kea
  has always had within the indigenous Hawaiian community. We are most
  fortunate to have the opportunity to conduct observations from this
  mountain.

  \facilities{IRTF (SpeX), ASAS-SN, Pan-STARRS\,1, ATLAS, Gaia, WISE,
    NEOWISE, IRSA, LCOGT (Spectral), NTT (SOFI)}

\end{acknowledgments}

\clearpage
%\bibliography{/users/mliu/tex/bibtex/mliu}
%\bibliographystyle{aasjournal}
%\bibliographystyle{apj}

%======================================================================%
\clearpage
%% use the figure environment and \plotone or \plottwo to include 

\clearpage
% ================================================================================ %
% Table: IRTF/SpeX spectra
% ================================================================================ %
\begin{deluxetable}{lccccccccccccc}
%\setlength{\tabcolsep}{0.035in}
%\rotate
%\tabletypesize{\small}
\tablecaption{IRTF/SpeX Spectroscopy Log \label{table:spex}}
\tablewidth{0pt}
\tablehead{
  \colhead{Date} &
  \colhead{\new{MJD}} &
  \colhead{$<$Airmass$>$} &
  \colhead{$T_{int}$} &
  \colhead{\new{A0V star}} &
  \multicolumn{4}{c}{Median S/N per pixel} &
  \colhead{\new{Spectral Type}} &
  \colhead{\new{$J-K$}} &
  \multicolumn{2}{c}{\new{Extinction}} \\
  \colhead{(UT)} &
  \colhead{} &
  \colhead{} &
  \colhead{(s)} &
  \colhead{} &
  \colhead{$Y$} &
  \colhead{$J$} &
  \colhead{$H$} &
  \colhead{$K$} &
  \colhead{} &
  \colhead{(mag)} &
  \colhead{$\Delta A_V$[mag]} &
  \colhead{\rchisq}
 }

\startdata     % DO NOT LEAVE THE NEXT LINE BLANK - Latex will crash
% SPEX_SUMMARIZE.PRO: Fri Dec 17 14:43:46 2021 mliu@sdhcp48.ifa.hawaii.edu 
% infile = "LIST.spectra"                                                                                                                                          File                                               N x Itime
2008-11-28\tablenotemark{a}       & 54798.6 & 1.52  &  1320.0  & HD 46680 &  21  &  43  &  41  &   45  &  M5.3$\pm$0.9 \intg  &  1.93 & \phs1.74  & 1.2    \\    % 2MASSJ0619-2903_0.5prism_2015dec24.fits            13_x_119.5
2015-12-24\tablenotemark{b}       & 57380.4 & 1.71  &  1554.1  & HD 41473 &  69  &  98  &  97  &  101  &  M5.8$\pm$0.9 \vlg   &  1.66 & \phs0.00 &  n/a    \\    % 2MASSJ06195260-2903592_0.5prism_2008nov28.fits     11_x_120.0
2020-11-27                        & 59180.4 & 1.52  &  1554.1  & HD 41473 &  52  &  71  &  71  &   65  &  M6.1$\pm$0.9 \vlg   &  1.77 & \phs0.71 &  1.7    \\    % 2MASS_J06195260-2903592_0.5prism_2020dec28.fits    13_x_59.8 
2020-12-28                        & 59211.4 & 1.52  &   777.0  & HD 56751 &  27  &  39  &  35  &   40  &  M6.6$\pm$0.9 \vlg   &  1.84 & \phs1.08 &  1.1    \\    % 2MASS_J06195260-2903592_0.5prism_2020dec31.fits     9_x_119.5
2020-12-31                        & 59214.4 & 2.12  &  1075.9  & HD 50917 &  39  &  53  &  51  &   57  &  M5.9$\pm$0.9 \vlg   &  1.86 & \phs1.18 &  1.3    \\    % 2MASS_J06195260-2903592_0.5prism_2020nov27.fits    13_x_119.5
2021-04-01                        & 59305.2 & 1.78  &  1793.2  & HD 50917 &  20  &  25  &  18  &   19  &  M6.4$\pm$0.9 \intg  &  1.56 &  $-$0.74 &  0.8    \\    % 2MASS_J06195260-2903592_0.5prism_2021apr01.fits    15_x_119.5
2021-04-03                        & 59307.2 & 1.68  &  2630.0  & HD 50917 &  48  &  60  &  49  &   54  &  M5.9$\pm$0.9 \vlg   &  1.61 &  $-$0.37 &  1.7    \\    % 2MASS_J06195260-2903592_0.5prism_2021apr03.fits    22_x_119.5
\new{2022-01-18}                  & 59597.4 & 1.60  &   777.0  & HD 43070 &  54  &  70  &  65  &   60  &  M5.9$\pm$0.9 \vlg   &  1.62 &  $-$0.32 &  1.6    \\    % 2MASS_J06195260-2903592_0.5prism_2022jan18.fits    13_x_59.8
\new{2022-02-23}                  & 59633.3 & 1.73  &   896.6  & HD 43070 &  35  &  46  &  39  &   40  &  M6.2$\pm$0.9 \vlg   &  1.62 &  $-$0.24 &  1.0    \\    % 2MASS_J06195260-2903592_0.8prism_2022feb23.fits    15_x_59.8
\enddata

\tablecomments{For each epoch's data, the MJD, $<$Airmass$>$, $T_{int}$,
  A0V star, and $(J-K)$ columns give the Modified Julian Date, average
  airmass, the total integration time, the telluric calibrator, and
  the synthesized $J-K$ color on the MKO photometry system
  \citep{2002PASP..114..169S}, respectively. A conservative uncertainty
  estimate for the synthesized $J-K$ color would be 0.05~mag
  \citep[e.g.][]{2012ApJS..201...19D, 2012ApJ...750..105R}. The
  Extinction columns give
  the best-fitting relative extinction (assuming $R=3.1$)
  and associated reduced \chisq\ of each epoch compared to the 2015
  epoch, which has the highest S/N.
  Negative values indicate those epochs's spectra were bluer than the
  2015 epoch. The formal errors on the fitted ${\Delta}A_V$ are
  negligible, though a conservative assumption on the overall flux calibration of
  the spectra would indicate an uncertainty of 0.3~mag
  (Section~\ref{sec:nearIR}).}

\tablenotetext{a}{From \citet{2013ApJ...772...79A}.}

\tablenotetext{b}{From \citet{2016ApJ...833...96L}.}

\end{deluxetable}

\end{document}